\documentclass{ws-ijmpd}
\usepackage{graphicx,epsf, epsfig, amssymb}
\usepackage[super, compress]{cite}
\usepackage[breaklinks]{hyperref}
\usepackage{url}
\usepackage{yfonts}
\usepackage{multirow}

\usepackage{tikz}
\usetikzlibrary{shapes.geometric, arrows}

\tikzstyle{startstop} = [rectangle, rounded corners, minimum width=3cm, minimum height=1cm,text centered, draw=black, fill=orange!15] 
\tikzstyle{paper} = [rectangle, minimum width=3cm, minimum height=1cm,text centered, draw=black, fill=blue!5] 
\tikzstyle{textbox} = [rectangle, rounded corners, minimum width=3cm, minimum height=1cm,text centered, draw=black, text width=4cm, align=center, fill=orange!20] 
\tikzstyle{process} = [rectangle, minimum width=3cm, minimum height=1cm, text centered, draw=black, fill=orange!30] 
\tikzstyle{invisible} = [rectangle, rounded corners, minimum width=3cm, minimum height=1cm, text centered, draw=yellow, fill=white] 
\tikzstyle{arrow} = [,->,>=latex] 
\tikzstyle{arrowd} = [dashed,->,>=latex] 

\usepackage{color,hyperref}
\definecolor{darkblue}{rgb}{0.0,0.0,0.3}
\definecolor{venetianred}{rgb}{0.78, 0.03, 0.08}
\definecolor{grey}{rgb}{0.25, 0.25, 0.28}
\definecolor{wenge}{rgb}{0.39, 0.33, 0.32}
\definecolor{bistre}{rgb}{0.24, 0.17, 0.12}
\definecolor{arsenic}{rgb}{0.23, 0.27, 0.29}
\definecolor{flame}{rgb}{0.89, 0.35, 0.13}
\definecolor{darkmidnightblue}{rgb}{0.0, 0.2, 0.4}
\hypersetup{colorlinks,breaklinks,
            linkcolor=darkmidnightblue,urlcolor=venetianred,
            anchorcolor=flame,citecolor=darkmidnightblue}

\def\f{\frac}
\def\ii{{\rm i}}
\def\radcurv{\mathfrak{r}} 

\newcommand{\gothr}{\textswab{r}}

\newcommand{\tn}{\textnormal}

\def\be{\begin{equation}}
\def\ee{\end{equation}}
\def\bea{\begin{eqnarray}}
\def\eea{\end{eqnarray}}
\newcommand{\beq}{\begin{eqnarray}}
\newcommand{\eeq}{\end{eqnarray}} 
\newcommand{\ba}{\begin{align}}
\newcommand{\ea}{\end{align}}
\begin{document}

\markboth{Berti, Cardoso, Crispino, Gualtieri, Herdeiro, Sperhake}
{Instructions for Typing Manuscripts (Paper's Title)}

%
\catchline{}{}{}{}{}
%

\title{Numerical Relativity and High Energy Physics: \\ Recent Developments}

\author{Emanuele Berti${}^{1,2}$, Vitor Cardoso${}^{2,3,4,5,1}$, Luis C. B. Crispino${}^4$, \\
        Leonardo Gualtieri${}^5$, Carlos Herdeiro${}^{6}$ and Ulrich Sperhake${}^{7,8,1}$}

\address{
${}^1\,$Department of Physics and Astronomy, The University of
Mississippi \\ 
University, MS 38677, USA\\[5pt]
${}^2\,$CENTRA, Departamento de F\'isica, Instituto Superior T\'ecnico,
Universidade de Lisboa\\
Avenida Rovisco Pais 1, 1049 Lisboa, Portugal\\[5pt]
${}^3\,$Perimeter Institute for Theoretical Physics, 31 Caroline Street North
Waterloo, Ontario N2L 2Y5, Canada\\[5pt]
${}^4\,$Faculdade de F\'{\i}sica, Universidade Federal do Par\'a, 66075-110, Bel\'em, Par\'a, Brazil
${}^5\,$Dipartimento di Fisica, ``Sapienza'' Universit\`a di Roma \& Sezione INFN Roma1, P.A. Moro 5, 00185, Roma, Italy\\[5pt]
${}^6\,$Departamento de F\'\i sica da Universidade de Aveiro and Center for Research in Mathematics and Applications (CIDMA), Campus de Santiago, 3810-183 Aveiro, Portugal\\[5pt]
${}^7\,$Department of Applied Mathematics and Theoretical Physics \\
University of Cambridge, Cambridge CB3 0WA, United Kingdom
\\[5pt]
${}^8\,$California Institute of Technology, Pasadena, CA 91125, USA \\[5pt]
}

\maketitle

\begin{history}
\received{Day Month Year}
\revised{Day Month Year}
\end{history}

\begin{abstract}
  We review recent progress in the application of numerical relativity
  techniques to astrophysics and high-energy physics. We focus on some
  developments that took place within the ``Numerical Relativity and
  High Energy Physics'' network, a Marie Curie IRSES action that we
  coordinated, namely: spin evolution in black hole binaries,
  high-energy black hole collisions, compact object solutions in
  scalar-tensor gravity, superradiant instabilities and hairy black
  hole solutions in Einstein's gravity coupled to fundamental fields,
  and the possibility to gain insight into these phenomena using
  analog gravity models.
\end{abstract}


\ccode{PACS numbers: 04.50.Kd, 04.70.-s, 04.70.Bw, 04.80.Cc}

\newpage

\tableofcontents

\newpage

\section{Introduction}	
 
The last global meeting of the Numerical Relativity and High Energy
Physics network -- a Marie Curie IRSES (International Research Staff
Exchange Scheme) partnership (2012-2015) funded by the European Union
and coordinated by the authors of this paper -- started in Bel\'em
(Brazil) on September 28, 2015. Everyone in attendance, as well as the
large majority of the scientific community, was unaware that a major
breakthrough in science had just taken place: precisely two weeks
earlier, the LIGO/Virgo collaboration observed the first
gravitational-wave (GW) signal from the merger of two black holes
(BHs)~\cite{Abbott:2016blz}.

The detection relied on decades of technological efforts to perform an
apparently impossible measurement, corresponding to displacements that
are $\sim 10^3$ times smaller than the atomic nucleus. The unambiguous
interpretation of the signal observed by Advanced LIGO as a BH binary
coalescence was also the result of a decades-long effort: it took over
40 years to numerically solve Einstein' equations of general
relativity (GR) and to understand the behavior BH binaries through
their inspiral, merger and ringdown.

A toolbox of powerful techniques became available after the tremendous
numerical relativity breakthrough that took place in
2005.\cite{Pretorius:2005gq,Baker:2005vv,Campanelli:2005dd} This
naturally led to a community effort looking for applications of these
tools beyond astrophysics~\cite{Cardoso:2012qm} and eventually to this
network, that looked at applications of numerical relativity both in
astrophysics~\cite{Berti:2015itd} and beyond~\cite{Cardoso:2014uka}.

This paper is a summary of some of the science produced within the
network. As such it is admittedly biased and incomplete, and it
certainly does not aim to be a comprehensive review of the impressive
developments in the area of numerical relativity and high-energy
physics that took place over the past few years. The plan of the paper
is as follows. We will start in Section~\ref{sec_2} by considering BH
collisions, first in astrophysics (paying particular attention to some
recent developments concerning spin dynamics), and then in the context
of fundamental and high-energy physics (with applications to large
extra dimension scenarios and to the gauge gravity duality). In
Section~\ref{sec_3} we will look at compact objects -- i.e., BHs and
neutron stars (NSs) -- in alternative theories of gravity, focusing on
models with scalar degrees of freedom in the gravitational sector:
tensor-(multi)-scalar theories, Horndeski gravity and
Eistein-dilaton-Gauss-Bonnet gravity.  In
Section~\ref{sec_superradiance} we will consider GR minimally coupled
to fundamental scalar and tensor fields and present some remarkable
results obtained in the last few years in these simple models,
including new types of numerical BH solutions that defied common
lore. The existence of these BHs with scalar or Proca hair is
intimately related with the complex phenomenon of superradiance, that
can occur for rotating and charged BHs. Numerical relativity
techniques have been (and will be) instrumental in probing the
dynamics of these objects.  Section~\ref{sec_analog} looks at many of
these phenomena (in particular those discussed in
Section~\ref{sec_superradiance}) from a different perspective: that of
analog gravity models. We close with some brief remarks.

\section{Black hole collisions: numerical and analytical studies}
\label{sec_2}

Collisions of BHs have been modeled using analytic and numerical
techniques for several decades. One of the main motivations throughout
this time has been the significance of merging BH binary systems as
one of the strongest sources for direct detection of GWs. The recent
breakthrough detection by Advanced LIGO of the event called GW150914
\cite{Abbott:2016blz} indeed observed the late stages of a BH binary
inspiral, including merger and ringdown. This event clearly marks a
revolution in our observational studies of the Universe. Astrophysical
BH binary mergers form a key motivation for the work reviewed here.
Additionally to this new era in gravitational astrophysics, many
developments in theoretical physics, particularly during the past two
decades, add substantial motivation to the modeling of BH collisions
from other angles\cite{Cardoso:2014uka,Sperhake:2014nra}, and make BHs
one of the centre-stage actors in contemporary physics.

A BH is the closest analog in GR to the concept of a point mass in
Newtonian physics, and spacetimes containing two BHs represent the
simplest version of the two-body problem in GR.  Unlike their
Newtonian counterparts, however, binary BH spacetimes have
substantially more complex dynamics: BHs have ``internal structure''
in the form of spin, and their interaction in a binary leads to GW
emission. Therefore it should come as no surprise that these
spacetimes cannot be described by exact solutions in closed analytic
form, analogous to the Keplerian orbits in Newtonian physics. For this
reason most theoretical modeling resorts to approximation methods,
such as post-Newtonian theory\cite{Blanchet:2006zz}, perturbation
theory\cite{Sasaki:2003xr} or the point-particle
approximation\cite{Poisson:2011nh}.  In alternative to these
approaches, which approximate the theory, numerical relativity
generates solutions to the full non-linear equations, approximating
them via some form of
discretization\cite{Alcubierre:2008,Bona:2009,Baumgarte2010}.  The
decades-long efforts of numerical relativity culminated in the 2005
breakthroughs performing the first evolutions of binary BHs through
inspiral, merger and ringdown
\cite{Pretorius:2005gq,Baker:2005vv,Campanelli:2005dd}; for a
historical perspective on this milestone see, e.g.,
Ref.~\refcite{Sperhake:2014wpa}.

The contemporary modeling of BH collisions in the context of
astrophysics, GW physics and high-energy physics relies on a
combination of all these analytic and numerical methods. The purpose
of this section is to review some of the most recent and exciting
developments.

\subsection{Astrophysical black holes and gravitational waves}
The modeling of BH binaries as astrophysical sources of GWs has mostly
focussed on systems in quasi-circular orbits because the emission of
GWs rapidly carries away excess angular momentum from the binary
\cite{Peters:1964zz}. By the time a binary has reached the frequency
window of GW detectors such as Advanced LIGO and Advanced Virgo, the
orbital eccentricity is very close to zero.  This efficient
elimination of eccentricity relies, of course, on the absence of any
significant interaction with matter or third bodies. The possible
effects of non-vanishing eccentricity have been investigated by
several analytical and numerical studies
\cite{Lincoln:1990ji,Mora:2003wt,Damour:2004bz,Hinder:2007qu,Sperhake:2007gu,Hinder:2008kv,Yunes:2009yz,Huerta:2013qb,Huerta:2014eca,Mishra:2015bqa}.
Quasicircular inspirals remain the most likely and best understood
scenario, so here we shall concentrate on this case.  We will also
focus on binaries in the framework of GR, but we note that there are
preliminary explorations of scalar radiation from BH binaries in
scalar-tensor theory, triggered either by non asymptotically flat
boundary conditions \cite{Berti:2013gfa} or by a non-vanishing
potential\cite{Healy:2011ef}.

The estimation of source parameters in GW observations employs a
method called {\em matched filtering} where the data stream is
compared with a catalog of theoretically predicted GW templates
\cite{Flanagan:1997kp}; for an application of this technique to hybrid
waveforms constructed out of numerical relativity and PN calculations
see for example Ref.~\refcite{Aasi:2014tra}. A main challenge for the
theoretical community is the generation of such template catalogs
covering with high accuracy the whole range of BH binary masses and
spins. Given the high computational cost of numerical relativity
simulations, this construction typically stitches together
post-Newtonian and numerical relativity
waveforms,\cite{Ajith:2012tt,Hinder:2013oqa,Khan:2015jqa} or employs
numerical simulations to calibrate free parameters in analytic
prescriptions such as the effective-one-body model
\cite{Pan:2013rra,Taracchini:2013rva,Damour:2014sva,Nagar:2015xqa}.

BH binaries with generic spins will undergo spin precession during
which the orbital plane changes orientation. The modeling of these
systems is significantly more involved than that of their
nonprecessing counterparts, but benefits enormously from the presence
of three distinctly different timescales. If we denote by $r$ the
separation of the two constituent BHs, these orbit around each other
on the {\em orbital} timescale $t_{\rm orb} \propto r^{3/2}$, while
the spin directions change on the {\em precession} timescale
$t_{\rm pre}\propto r^{5/2}$, and the emission of GWs reduces the
separation $r$ on the {\em radiation reaction} timescale
$t_{\rm RR}\propto r^4$. At sufficiently large separation $r$, this
implies the hierarchy $t_{\rm orb} \ll t_{\rm pre} \ll t_{\rm RR}$.
The first inequality has been used to derive {\em orbit-averaged}
evolution equations for the individual spin vectors $\boldsymbol{S}_i$
$(i=1,\,2)$ of the from
$\dot{\boldsymbol{S}}_i = \boldsymbol{\Omega}_i \times
\boldsymbol{S}_i$,
where the precession frequency depends on the orbital angular momentum
$\boldsymbol{L}$ and the $\boldsymbol{S}_i$, but not on the separation
vector
$\boldsymbol{r}$~\cite{Apostolatos:1994mx,Kidder:1995zr,Arun:2008kb}.
This {\em quasiadiabatic} approach has been combined with some
additional simplifications for the precession dynamics in order to
construct template banks for precessing binaries.  These techniques
include a single {\em effective} spin model, modifications to the {\em
  stationary-phase approximation}, or the use of nonprecessing
templates modulated through an {\em effective precession
  parameter}\cite{Apostolatos:1995pj,Hannam:2013oca,Klein:2013qda,Klein:2014bua,2014PhRvD..89d4021L,Schmidt:2014iyl}. Orbit-averaged
PN calculations have also been employed in the discovery of spin-orbit
resonances\cite{Schnittman:2004vq} and for predictions of the final
spins and recoil in BH binary mergers\cite{Kesden:2010yp,Kesden:2010ji,Berti:2012zp}.

The success of the orbit-averaging procedure relies heavily on the
analytic solutions for Keplerian orbits that are employed in the
averaging over the orbital timescale. Until recently, no analogous
analytic solution was known for the precession equations, so that the
second inequality of the above hierarchy,
$t_{\rm orb} \ll t_{\rm RR}$, has not been brought to the same level
of fruition. This picture changed with the identification of analytic
solutions on the precessional
timescale\cite{Kesden:2014sla,Gerosa:2015tea}. Consider for this
purpose a BH binary with orbital angular momentum $\boldsymbol{L}$,
individual masses $m_i$ and spin vectors $\boldsymbol{S}_i,~i=1,\,2$
and mass ratio $q=m_2/m_1\le 1$. For fixed mass ratio, the system is
described by nine parameters, three each for $\boldsymbol{S}_1$,
$\boldsymbol{S}_2$ and $\boldsymbol{L}$. Conservation of the spin
magnitudes $S_i$ reduces this number to seven. On the precession
timescale, the total angular momentum
$\boldsymbol{J}\equiv
\boldsymbol{S}_1+\boldsymbol{S}_2+\boldsymbol{L}$
as well as the magnitude $L$ are also conserved at the PN orders
considered here, leaving three numbers to determine the state of the
binary.  A convenient choice for these variables is given by the
angles $\theta_i$ between the individual spins and the orbital angular
momentum vector and the angle $\Delta \Phi$ between the projections of
the individual spins onto the orbital plane: cf. e.g. Fig.~1 in
Ref.~\refcite{Gerosa:2015tea}. One further variable can be eliminated
through a convenient choice of a non-inertial frame. Finally, the
projected effective spin defined by\cite{Damour:2001tu,Racine:2008qv}
\begin{equation}
  \xi \equiv (m_1+m_2)^{-2}[(1+q)\boldsymbol{S}_1
        +(1+q^{-1}) \boldsymbol{S}_2]\cdot
        \boldsymbol{L}/L\,,
\end{equation}
is conserved by the orbit-averaged spin-precession equations at 2PN
order, and even under radiation reaction at 2.5PN order. Spin precession
at this order is therefore described in terms of a single evolution
variable, conveniently chosen to be the magnitude of the total spin
$S\equiv |\boldsymbol{S}_1+\boldsymbol{S}_2|$.

For a BH binary with specified parameters and separation,
i.e.~fixed values $m_i,\,S_i,\,L,\,J,\,\xi$, the precession is described
completely in terms of the variable $S$.
The set of physically allowed systems
can then be represented as the area inside a closed loop constructed
from two ``effective potentials'' $\xi_{\pm}(S)$ in the $(S,\xi)$ plane.
The functions $\xi_{\pm}(S)$ are determined by the physical constraints
on the spin and angular momenta
\begin{eqnarray}
  &S_{\rm min} = \text{max}(|J-L|,\,|S_1-S_2|)\,,~~~
  & S_{\rm max} = \text{min}(J+L,\,S_1+S_2)\,, \\
  &J_{\rm min} = \text{max}(0,\,L-S_1-S_2,\,|S_1-S_2|-L)\,,~~~
  & J_{\rm max} = L+S_1+S_2\,,
\end{eqnarray}
and are given in closed analytic form by Eq.~(14) in
Ref.~\refcite{Gerosa:2015tea}.  For a given value of $\xi$ inside the
range compatible with these constraints, this implies that $S$
oscillates between two extrema $S_{\pm}$, where
$S_{\rm min} \le S_- \le S_+\le S_{\rm max}$\footnote{The resonance
  configurations of Ref.~\refcite{Schnittman:2004vq} correspond to the maximal
  and minimal allowed values of $\xi$ in this area, $\xi_{\rm max}$
  and $\xi_{\rm min}$, at which $S_-=S_+$ and, hence, $S$ remains
  constant on the precession timescale; cf.~Fig.~2 in
  Ref.~\refcite{Gerosa:2015tea}.}.  All remaining variables of the
binary can be obtained from $S$ through Eqs.~(20) in
Ref.~\refcite{Gerosa:2015tea} for $\theta_1$, $\theta_2$ and
$\Delta \Phi$ which, in turn, determine all other physical variables.

A particularly intriguing consequence of this formulation of the spin
precession dynamics is that all binaries fall into one of three
morphologies, which are best characterized by the behavior of the
angle $\Delta \Phi$ on the precession time scale. As the variable $S$
oscillates inside its allowed range, $\Delta \Phi$ either (i) librates
around $0$, (ii) librates around $\pi$, or (iii) circulates through
the entire range $\Delta \Phi \in [-\pi,\pi]$. As the binary inspirals
on the much larger radiation reaction timescale, the orbital and total
angular momentum evolve, and the binary may undergo phase transitions
between these morphologies. The inspiral of the binary under GW
emission can be modeled in a remarkably efficient manner at 1PN order
if we express the binary separation $r$ in terms of the orbital
angular momentum $L$ given by the Newtonian expression
$L=\eta(rM^3)^{1/2}$, where $M\equiv m_1+m_2$ and
$\eta \equiv m_1 m_2/(m_1+m_2)$ is the symmetric mass ratio parameter.
The evolution of the total angular momentum $J$ averaged over a
precession cycle is then given by\cite{Gerosa:2015tea}
\begin{equation}
  \left\langle \frac{dJ}{dL} \right\rangle
        = \frac{1}{2LJ}(J^2-L^2- \langle S^2 \rangle_{\rm pre})\,,
  \label{eq:dJdL}
\end{equation}
where
\begin{eqnarray}
  &&\langle S^2 \rangle = \frac{2}{\tau} \int_{S_-}^{S_+}
        S^2 \frac{dS}{|dS/dt|}\,,~~~~~~~~~~~~~~
  \tau \equiv 2\int_{S_-}^{S_+} \frac{dS}{|dS/dt|}\,, \\
  && \frac{dS}{dr} = -\frac{3(1-q^2)}{2q}\,
        \frac{S_1 S_2}{S}\,
        \frac{(\eta^2 M^3)^3}{L^5}
        \left(1-\frac{\eta M^2 \xi}{L} \right)
        \sin \theta_1\,\sin\theta_2\,\sin \Delta \Phi\,.
\end{eqnarray}
The evolution of precessing BH binaries is thus modeled in terms of a
single ordinary differential equation (\ref{eq:dJdL}) which, thanks to
the precession-averaging procedure, can furthermore be solved
numerically using much larger timesteps than possible in a formulation
using only orbit-averaged variables. By suitably compactifying the
variables involved, accurate numerical evolutions from infinite
separations become possible at drastically reduced computational
cost. The formalism truly bridges the gap between astrophysical BH
separations and the regime close to merger, where numerical relativity
predictions for BH kicks are valid. Further applications of the
formalism identified a precessional instability of binaries where the
spin of the more (less) massive BH is (anti) aligned with the orbital
angular momentum \cite{Gerosa:2015hba}, and highlighted how the
precessional morphology may carry a memory of the astrophysical
processes that formed the binary\cite{Gerosa:2013laa,Gerosa:2015tea}.
Preliminary studies of the potential of present and future GW
detectors to determine the morphologies in BH observations are
encouraging, except for highly symmetric binaries, where precessional
effects are suppressed\cite{Gerosa:2014kta,Trifiro:2015zda}.

\subsection{Black hole collisions, fundamental and high-energy physics}

Even 100 years after its publication, GR still confronts us with some
of the most important questions in contemporary physics.  As described
below in Sec.~\ref{sec_superradiance}, astrophysical and cosmological
observations suggest the presence of an enigmatic dark sector which
appears to dominate the gravitational dynamics of much of our
universe. At an even more fundamental level, GR predicts the limits of
its own range of validity. Seminal work by Hawking and Penrose
\cite{Penrose:1964wq,Hawking:1967ju,Hawking:1969sw} demonstrated that
gravitational collapse in the framework of GR leads to singularities
under generic initial conditions.  The appearance of infinities in the
theory is expected to be cured by a future quantum theory of
gravitation. Quite remarkably, however, relativity appears to have a
built-in protection mechanism against the potentially fatal
consequences of spacetime singularities; according to Penrose's {\em
  cosmic censorship conjecture}, \cite{Wald:1997wa} the singularities
do not appear in {\em naked} form for physically realistic, generic
initial data, but instead are cloaked inside horizons which causally
disconnect the exterior spacetime from being influenced by the
singularity. According to Thorne's {\em hoop conjecture},
\cite{Thorne:1972ji} such horizons should furthermore form (in $D=4$
spacetime dimensions) whenever a physical system of mass $M$ gets
compacted inside a region with circumference $\lesssim 2\pi R_s$,
where $R_s=2M$ is the Schwarzschild radius associated with $M$. The
conjecture has also been generalized to higher
dimensions\cite{Ida:2002hg}.

A particularly intriguing consequence arising from the hoop conjecture
is the possibility of BH formation in proton-proton collisions
at colliders such as the LHC, or in cosmic-ray showers hitting the
Earth's atmosphere. In the trans-Planckian regime, where the colliding
partons can be approximated as classical particles, the hoop conjecture
predicts formation of a BH if the boost parameter $\gamma$ satisfies
$\gamma \gtrsim c^4 R/(4Gm_0)$, i.e.~if two particles of rest mass $m_0$
with center-of-mass energy $M=2\gamma m_0$ get compacted inside a volume
of radius $\sim R$.
Taking the radius to be given by the de Broglie wavelength
$hc/M$ associated with the center-of-mass energy, the condition
for BH formation becomes
\cite{Choptuik:2009ww} (up to factors of order unity)
$M \gtrsim E_{\rm P} = \sqrt{\hbar c^5/G}$, i.e.~the center-of-mass
energy of the collision must exceed the Planck energy.
At the four-dimensional standard-model value $E_p\sim 10^{19}~{\rm GeV}$,
experimental tests of BH formation are clearly out of the range
of present and forseeable colliders. The so-called TeV gravity
scenarios involving large or warped extra dimensions
\cite{Antoniadis:1998ig,ArkaniHamed:1998rs,Randall:1999ee,Randall:1999vf},
however, provide an appealing explanation of the {\em hierarchy
  problem} of physics and may lower the effective Planck energy to
values as low as the TeV regime, which would allow for the possibility
to form BHs in particle collisions at the
LHC\cite{Dimopoulos:2001hw,Giddings:2001bu}.  Simulations of BH
collisions can provide important information about the cross section
and energy loss through GWs which form key input for the Monte-Carlo
generators employed in the analysis of experimental
data~\cite{Dimopoulos:2001hw,Frost:2009cf}.

Yet another rich area of applying BH studies has emerged in the
context of the gauge-gravity duality, often also referred to as
the AdS/CFT correspondence,\cite{Maldacena:1997re,Gubser:1998bc,Witten:1998qj}
which states the equivalence between string theory
in asymptotically AdS spacetimes (times a compact space) and conformal
field theories living on the AdS boundary. The duality provides a
new approach to the (notoriously difficult) modeling of physical
systems in strongly coupled gauge theories in terms of classical
spacetimes, often involving BHs, that are one dimension higher
and asymptote to AdS at infinite radius.

In the following we will review some of the recent developments in BH
modeling in the context of these topics, but we note that there are
several more comprehensive reviews on these
subjects~\cite{Pretorius:2007nq,Sperhake:2011xk,Cardoso:2012qm,Cardoso:2014uka}.

The hoop conjecture has been tested in the context of high-speed
collisions of scalar-field \cite{Choptuik:2009ww} and fluid-ball
configurations\cite{East:2012mb,Rezzolla:2012nr}.  In all these
simulations, BH formation is observed at high velocities, consistent
with the prediction of the hoop conjecture. Combined with the fact
that in high-energy collisions most of the center-of-mass energy is
present in the form of kinetic energy, the hoop conjecture supports
the idea that parton-parton collisions can be well modeled by
colliding BHs, the GR analog of point particles.  High-energy
collisions have been studied most comprehensively in $D=4$ spacetime
dimensions and revealed a number of intriguing features.  Numerical
simulations of equal-mass, non-spinning BHs colliding head-on
predict\cite{Sperhake:2008ga} that in the ultrarelativistic limit a
fraction of $14\pm 3\%$ (recently confirmed and refined to
$13\pm 1\%$ by the RIT group\cite{Healy:2015mla}) of the total energy
is radiated in GWs. This value is about half of Penrose's upper
limit,\cite{Penrose1974,Eardley:2002re} and in good agreement with the
value of $16.4\%$ obtained in second-order perturbative calculations
on a background composed of two superposed Aichelburg-Sexl shock waves
\cite{D'Eath:1976ri,D'Eath:1992hb,D'Eath:1992hd,Herdeiro:2011ck,Coelho:2012sya,Coelho:2012sy,Coelho:2013zs,Coelho:2014gma,Coelho:2015qaa}.

In grazing collisions the BHs are allowed to approach each other with
a non-zero impact parameter $b$, and the outcome of the collision
depends on whether this parameter exceeds a threshold value or
not. This {\em scattering threshold} $b_{\rm scat}$ separates
configurations that result in the formation of a single BH
($b<b_{\rm scat}$) or in the constituents scattering off each other to
infinity ($b>b_{\rm scat}$), and was shown to be approximately given
as a function of the collision speed $v$ and the center-of-mass energy
$Mc^2$ by the remarkably simple formula ~\cite{Shibata:2008rq}
\begin{equation}
  b_{\rm scat} \approx 2.5 \frac{GM}{cv}\,.
\end{equation}
Numerical simulations~\cite{Sperhake:2009jz} furthermore identified
the presence of zoom-whirl orbits\cite{Pretorius:2007jn} in a regime
where $b$ is close to a critical value $b_* \lesssim b_{\rm scat}$.
The energy released in GWs in these grazing collisions can be
enormous, exceeding $35\%$ of the total energy. Extrapolation to the
speed of light, however, demonstrates that the maximum energy
saturates at about half of the total (i.e.~kinetic)
energy\cite{Sperhake:2012me}.  The remaining kinetic energy, instead,
ends up as rest mass, either in the single BH resulting from merger or
in the two constituents in scattering configurations.  Simulations of
rotating BHs \cite{Sperhake:2012me} also demonstrate that the impact
of the spin on the collision dynamics is washed out in the limit
$v\rightarrow c$.  We find here another confirmation of the ``matter
does not matter'' conjecture already encountered in scalar-field and
fluid-ball
collisions\cite{Choptuik:2009ww,East:2012mb,Rezzolla:2012nr}:
ultrarelativistic collisions are dominated by the kinetic energy, so
that the internal structure of the colliding objects becomes
irrelevant for the collision process. Further evidence for this
conjecture has recently been obtained in numerical studies of
unequal-mass BH collisions.  Head-on collisions of this type emit
$\sim 13\%$ of the total mass in the form of GWs in the
ultrarelativistic limit, in excellent agreement with the equal-mass
result mentioned above\cite{Sperhake:2015siy}.

It remains to be seen whether this picture remains intact as electric
charge is added to the colliding particles. Collisions of electrically
charged BHs have so far considered only the low-velocity regime, and
revealed qualitatively similar dynamics as for the case of neutral
BHs\cite{Zilhao:2012gp}.  As intuitively expected, however, the
collision is slowed down in the case of equal electric charges,
reducing the energy radiated in electromagnetic and GWs as the
charge-to-mass ratio approaches the critical value $Q/M=1$.
Conversely, the collision is accelerated by the additional attractive
force between the BHs if they carry opposite charges, which increases
the GW radiation by a factor of $\sim 2.7$ as the charge-to-mass ratio
$|Q|/M$ increases from 0 to 0.99 ~\cite{Zilhao:2013nda}. The
electromagnetic radiation becomes dominant in these collisions at
$|Q|/M\sim 0.37$ and exceeds its GW counterpart by a factor $\sim 5.8$
when $|Q|/M=0.99$.

The high-energy collision of particles has also attracted considerable
interest in a more astrophysical context through the {\em collisional
  Penrose} process\cite{Piran:1975}. Particle collisions near rapidly
rotating BHs could in principle lead to arbitrarily large
center-of-mass energies\cite{Banados:2009pr}, but there are several
caveats on the astrophysical viability of this
process\cite{Berti:2009bk,Jacobson:2009zg,McWilliams:2012nx}. The
significance of such collisions is limited, in particular, by the
redshift experienced by particles escaping from the near-horizon area
to observers far away from the
source\cite{Piran:1977dm,Harada:2012ap}. An interesting possibility is
that one of the colliding particles could have outgoing radial
momentum: this can happen either because the particle reaches a
turning point in the orbit\cite{Schnittman:2014zsa}, or by allowing at
least one outgoing particle to be generated close to the BH via
previous collisions\cite{Berti:2014lva}.  In both cases the efficiency
of the process (i.e., the ratio of the escaping particle's energy to
the sum of the pre-collision particles' energies) can reach values as
large as $\sim 13.9$\cite{Ogasawara:2015umo,Leiderschneider:2015kwa}.

Collisions of BHs in $D>4$ spacetime dimensions are not as well
understood as the $D=4$ case, mostly because of difficulties arising
in the numerical stability of the simulations. The most extensive
exploration of BH collisions in $D=5$ was able to determine the
scattering threshold in grazing collisions at velocities up to
$v\approx 0.6~c$\cite{Okawa:2011fv}.  This study used the so-called
{\em modified Cartoon
  method}\cite{Alcubierre:1999ab,Pretorius:2004jg,Yoshino:2011zz,Yoshino:2011zza,Cook:2016soy}
and identified regions of exceptionally high curvature above the
Planck regime that are not hidden inside a BH horizon. These regions
with curvature radius below the Planck length are realized during the
close encounter of two BHs in scattering configurations.  The emission
of GWs in $D=5$ has been analyzed in head-on collisions of
non-spinning BHs starting from rest\cite{Witek:2010xi,Witek:2010az}.
It predicts $E_{\rm rad}/M = (0.089 \pm 0.006)\%$ for equal masses
and a monotonic decrease, in good agreement with point particle
approximations, as the mass ratio is lowered from $q=1$ to smaller
values. The numerical method of this work is based on a dimensional
reduction by isometry\cite{Zilhao:2010sr}, analogous to the Geroch
decomposition \cite{Geroch:1970nt}.  Results from the two different
codes have been compared in $D=5$, demonstrating excellent agreement
~\cite{Witek:2014mha}. This work also provided the first estimate in
$D=6$, where equal-mass head-on collisions yield
$E_{\rm rad}/M = (0.081 \pm 0.004)\%$.  All these studies assume
rotational symmetry in planes involving the extra dimensions such that
the effective computational domain remains three dimensional.  These
symmetry assumptions still accomodate most of the scenarios relevant
in the context of testing TeV gravity or fundamental properties of BH
spacetimes.

Perturbative calculations based on superposed shock waves have also
been extended to $D\ge 5$ dimensions, including non-zero impact
parameters\cite{Eardley:2002re,Yoshino:2002tx,Yoshino:2005hi}.  These
calculations predict a significant increase of the threshold impact
parameter for formation of a common apparent horizon relative to the
four-dimensional case; see, in particular, Table II in
Ref.~\refcite{Yoshino:2005hi}. Extension of the work by d'Eath and
Payne for the head-on case to $D\ge 5$ resulted in a remarkably simple
expression at first perturbative order for the energy fraction
radiated in GWs \cite{Herdeiro:2011ck,Coelho:2012sya}:
$E_{\rm rad}/M = 1/2-1/D$.  This result, originally obtained in a
numerical study, has more recently been confirmed analytically to be
exact at first order \cite{Coelho:2014gma}.

The modeling of shock-wave and BH collisions in asymptotically AdS
spacetimes is significantly complicated by the active role and the
complex structure of the outer boundary: see
Refs.~\refcite{Bantilan:2012vu,Chesler:2013lia} for a discussion of
numerical methods. Over the past decade, substantial progress has been
made in overcoming these difficulties, achieving the first collision
of BHs in an asymptotically AdS spacetime\cite{Bantilan:2014sra} (see
also Ref.~\refcite{Witek:2010qc} for an earlier toy model).  Numerical
simulations of shock waves and BHs have clearly demonstrated their
capacity for obtaining new insight into the strongly coupled regime of
gauge theories. These studies have addressed in particular the
thermalization of quark-gluon plasma in the heavy-ion collisions
performed for example at the Brookhaven RHIC collider. Estimates for
the thermalization time obtained through the AdS/CFT correspondence
are $\sim 0.35~{\rm fm}/c$, in good agreement with experimental
data~\cite{Chesler:2008hg,Chesler:2010bi,Heller:2012je}. As in the
asymptotically flat case, point-particle and perturbative calculations
provide a convenient tool complementing full numerical relativity
studies \cite{Heller:2012km,Cardoso:2013vpa}. For a more extended
discussion of BH and shock wave collisions in AdS, we refer the reader
to Ref.~\refcite{Cardoso:2014uka}.

Finally, we remark that BH collisions in asymptotically de Sitter
spacetimes have been considered as tests of the cosmic censorship
conjecture. The numerical simulations support the
conjecture.\cite{Zilhao:2012bb}

\section{Compact objects in modified theories of gravity}
\label{sec_3}

In this section we will discuss isolated compact objects (BHs and NSs)
in modified theories of gravity.  We will focus on one of the most
natural and best studied extensions of GR: {\it scalar-tensor
  gravity}, in which one or more scalar degrees of freedom are
included in the gravitational sector through a nonminimal
coupling~\cite{Damour:1992we,Chiba:1997ms,2003sttg.book.....F,Faraoni:2004pi,Herdeiro:2015waa,Sotiriou:2015lxa}.

We will discuss compact objects in these theories at increasing
levels of complexity, starting from the ``standard'' Bergmann-Wagoner
formulation (Section~\ref{sec:BW}) and then considering extensions to
multiple scalar fields (Section~\ref{sec:TMS}), Horndeski gravity
(Section~\ref{sec:Horndeski}) and Einstein-dilaton-Gauss-Bonnet
gravity (Section~\ref{sec:EdGB}).
%

\subsection{``Bergmann-Wagoner'' scalar-tensor theories}
\label{sec:BW}

The most general action of scalar-tensor gravity, at most quadratic in
derivatives of the fields and with one scalar field, was studied by
Bergmann and Wagoner~\cite{Bergmann:1968ve,Wagoner:1970vr}.  The
action of this theory can be written (with an appropriate field
redefinition) as:
\begin{equation}
  S=\frac{1}{16\pi}\int d^4x\sqrt{-g}\left[\phi R-\frac{\omega(\phi)}{\phi}g^{\mu\nu}
    \partial_\mu\phi\, \partial_\nu\phi
    -U(\phi)\right]+S_M[\Psi,g_{\mu\nu}]\,,\label{STactionJ}
\end{equation}
where $U(\phi)$ and $\omega(\phi)$ are arbitrary functions of the
scalar field $\phi$, and $S_M$ is the action of the matter fields
$\Psi$. When $U(\phi)=0$ and $\omega(\phi)=\omega_{BD}$ is constant,
the theory reduces to (Jordan-Fierz-)Brans-Dicke
gravity~\cite{Jordan:1959eg,Fierz:1956zz,Brans:1961sx}.

The Bergmann-Wagoner theory \eqref{STactionJ} can be expressed in a
different form through a scalar field redefinition
$\varphi=\varphi(\phi)$ and a conformal transformation of the metric
$g_{\mu\nu}\rightarrow g^\star_{\mu\nu}=A^{-2}(\varphi)
g_{\mu\nu}$.
In particular, fixing $A(\varphi)=\phi^{-1/2}$, the {\it Jordan-frame}
action~(\ref{STactionJ}) transforms into the {\it Einstein-frame}
action
\begin{equation}
  S=\frac{1}{16\pi}\int d^4x\sqrt{-g^\star}\left[R^\star-2g^{\star\mu\nu}
    \partial_\mu\varphi\, \partial_\nu\varphi
    -V(\varphi)\right]+S_M[\Psi,A^2(\varphi)g^\star_{\mu\nu}]\,,\label{STactionE}
\end{equation}
where $g^\star$ and $R^\star$ are the determinant and Ricci scalar of
$g_{\mu\nu}^{\star}$, respectively, and the potential
$V(\varphi) \equiv A^4(\varphi)U(\phi(\varphi))$.  The price paid for
the minimal coupling of the scalar field in the gravitational sector
is the nontrivial coupling in the matter sector of the action:
particle masses and fundamental constants depend on the scalar field.

The actions~\eqref{STactionJ} and \eqref{STactionE} are just different
representations of the {\em same}
theory~\cite{Flanagan:2004bz,Sotiriou:2007zu}, so it is legitimate
(and customary) to choose the conformal frame in which calculations
are simpler. For instance, in vacuum the Einstein-frame action
\eqref{STactionE} formally reduces to the GR action minimally coupled
with a scalar field. It may then be necessary to change the conformal
frame when extracting physically meaningful statements (since the
scalar field is minimally coupled to matter in the Jordan frame, test
particles follow geodesics of the {\em Jordan-frame} metric, not of
the Einstein-frame metric).  The relation between Jordan-frame and
Einstein-frame quantites is simply $\phi=A^{-2}(\varphi)$,
$3+2\omega(\phi)=\alpha(\varphi)^{-2}$, where
$\alpha(\varphi)\equiv d(\ln
A(\varphi))/d\varphi$~\cite{Will:2014xja}.

Many phenomenological studies neglect the scalar potential, setting
$U(\phi)=0$ or $V(\varphi)=0$.\footnote{This approximation corresponds
  to neglecting the cosmological term, the mass of the scalar field
  and possible self-interactions. In an asymptotically flat spacetime
  the scalar field tends to a constant $\phi_0$ at spatial infinity,
  corresponding to a minimum of $U(\phi)$. Taylor expanding $U(\phi)$
  around $\phi_0$ yields, at the lowest orders, a cosmological
  constant and a mass term for the scalar
  field~\cite{Wagoner:1970vr,Alsing:2011er}.}  If the potential
vanishes, the theory is characterized by a single function
$\alpha(\varphi)$. The expansion of this function around the
asymptotic value $\varphi_0$ can be written in the form
\begin{equation}
\alpha(\varphi)=\alpha_0+\beta_0(\varphi-\varphi_0)+\dots\label{DEalphabeta}
\end{equation}
The choice $\alpha(\varphi)=\alpha_0=$constant (i.e.,
$\omega(\phi)=$constant) corresponds to Brans-Dicke theory. A more
general formulation, proposed by Damour and Esposito-Far\`ese, is
parametrized by $\alpha_0$ and
$\beta_0$~\cite{Damour:1993hw,Damour:1996ke}. Another simple variant
is massive Brans-Dicke theory, in which $\alpha(\varphi)$ is constant,
but the potential is nonvanishing and has the form
$U(\phi)=\f{1}{2}U''(\phi_0)(\phi-\phi_0)^2$, so that the scalar field
has a mass $m_s^2\sim U''(\phi_0)$. Since the scalar field $\varphi$
in the action (\ref{STactionE}) is dimensionless, the function
$\alpha(\varphi)$ and the constants $\alpha_0$, $\beta_0$ are
dimensionless as well.

In the Einstein frame, the field equations are
\begin{subequations}
\label{ST:Einstein}  
\begin{align}
 G^\star_{\mu \nu} &=2\left(\partial_\mu\varphi\partial_\nu\varphi-
 \frac{1}{2}g^\star_{\mu\nu}\partial_\sigma\varphi\partial^\sigma\varphi\right)-
 \frac{1}{2}g^\star_{\mu\nu}V(\varphi)+8\pi T^\star_{\mu\nu}\,,\label{eq:tensoreqnE}  \\
\Box_{g^\star} \varphi &=-4\pi\alpha(\varphi)T^\star+\frac{1}{4}\frac{dV}{d\varphi}\,,\label{eq:scalareqnE}
\end{align}
\end{subequations}
where
\begin{equation}
T^{\star\,\mu\nu}=-\f{2}{\sqrt{-g}} \f{\delta
S_M(\Psi,A^2g^\star_{\mu\nu})}{\delta g^\star_{\mu\nu}}
\end{equation}
is the Einstein-frame stress-energy tensor of matter fields, and
$T^\star=g^{\star\,\mu\nu}T^\star_{\mu\nu}$ is its trace. Eq.~\eqref{eq:scalareqnE}
shows that $\alpha(\varphi)$ determines the strength of the coupling
of the scalar fields to matter~\cite{Damour:1992we,Damour:1995kt}.

Astrophysical observations set bounds on the parameter space of
scalar-tensor theories. In the case of Brans-Dicke theory, the best
observational bound ($\alpha_0<3.5\times 10^{-3}$) comes from the
Cassini measurement of the Shapiro time delay~\cite{Will:2005va}. An
interesting feature of scalar-tensor gravity is the prediction of
characteristic physical phenomena which do not occur in GR. Even
though we know from observations that $\alpha_0\ll1$ and that GR
deviations are generally small, these phenomena may lead to observable
consequences. The best known example is the fact that compact binary
systems in scalar-tensor gravity emit dipolar gravitational
radiation~\cite{Eardley:1975,Will:1989sk}.  Dipolar gravitational
radiation is ``pre-Newtonian,'' i.e.~it occurs at lower PN order than
quadrupole radiation, and it does not exist in GR.
In the more general case with $\beta_0\neq0$, the phenomenon of
spontaneous scalarization (described below) can lead in principle to
macroscopic modifications in the structure of NSs, significantly
affecting the amount of dipolar radiation emitted by a binary
system. Therefore the best constraints in the $(\alpha_0,\beta_0)$
plane come from observations of NS-NS and NS-WD binary
systems~\cite{Freire:2012mg}.  Observations of compact binary systems
also constrain massive Brans-Dicke theory, leading to exclusion
regions in the $(\alpha_0,m_s)$ plane~\cite{Alsing:2011er}.

\subsubsection{Spontaneous scalarization in compact stars}

An interesting feature of scalar-tensor theories is the existence of
nonperturbative NS solutions in which the scalar field amplitude is
finite even for $\alpha_0\ll1$: this phenomenon, known as {\it
  spontaneous scalarization}~\cite{Damour:1993hw,Damour:1996ke}, may
significantly affect the mass and radius of a NS, and therefore the
orbital motion of a compact binary system, even far from coalescence.
A simple way to illustrate the principle behind spontaneous
scalarization is by taking the limit in which the scalar field
$\varphi$ is a small perturbation around a GR
solution~\cite{Yunes:2011aa}.  Expanding around the constant value
$\varphi_{0}$ to first order in
$\hat\varphi\equiv\varphi-\varphi_{0}\ll1$, the field equations in the
Einstein frame~(\ref{ST:Einstein}) read
\begin{subequations}
\begin{align}
  &G_{\mu\nu}^\star=8\pi T_{\mu\nu}^\star
  \,, \label{Einsteinlin} \\
&\square^\star\hat\varphi=-4\pi \alpha_0 T^\star-4\pi \beta_0\hat\varphi T^\star\,. \label{KGlin}
\end{align}
\end{subequations}
Here we have assumed analyticity around $\varphi\sim \varphi_{0}$ and
we have used Eq.~\eqref{DEalphabeta}.
It is clear from Eq.~\eqref{KGlin} that $\alpha_0$ controls the
effective coupling between the scalar and matter. Various
observations, such as weak-gravity constraints and tests of violations
of the strong equivalence principle, require $\alpha_0$ to be
negligibly small when the scalar tends to its asymptotic
value~\cite{Damour:1998jk,Damour:1996ke,Freire:2012mg}. This implies
that a configuration in which the scalar $\varphi\approx\varphi_0$ and
$\alpha_0\approx 0$ should be at least an approximate solution in most
viable scalar-tensor theories.  With $\alpha_0=0$, any background GR
solution solves the field equations above at first order in the scalar
field. At this order, the Klein-Gordon equation reads
\be
\left[\square^\star-\mu_s^2(x^\nu  )\right]\hat\varphi=0\,,\qquad \mu_s^2(x^\nu)\equiv -4\pi\beta_0 T^\star\,. \label{effectivemass}
\ee
Thus, the coupling of the scalar field to matter is equivalent to an
effective $x^\nu$-dependent mass. Depending on the sign of
$\beta_0 T^\star$, the effective mass squared can be negative. Because
typically\footnote{Some nuclear equations of state (EOSs) allow for
  positive $T^\star$ in the NS interior, with potentially interesting
  phenomenological
  implications\cite{Mendes:2014ufa,Palenzuela:2015ima}.}
$-T^\star\approx \rho^\star>0$ , this happens when $\beta_0<0$.  When
$\mu_s^2<0$ in a sufficiently large region inside the NS, scalar
perturbations of a GR equilibrium solution develop a tachyonic
instability associated with an exponentially growing mode, which
causes the growth of scalar hair in a process similar to
ferromagnetism~\cite{Damour:1993hw,Damour:1996ke}.

Spherically symmetric NSs develop spontaneous scalarization for
$\beta_0\lesssim-4.35$~\cite{Harada:1998}. Detailed investigations of
stellar structure~\cite{Damour:1996ke,Salgado:1998sg}, numerical
simulations of
collapse~\cite{Shibata:1994qd,Harada:1996wt,Novak:1997hw,Gerosa:2016fri}
and stability studies~\cite{Harada:1997mr,Harada:1998} confirmed that
spontaneously scalarized configurations would indeed be the end-state
of stellar collapse in these theories.  This is subject, however, to
the collapse process reaching a sufficient level of
compactness. Recent simulations of supernova core collapse identified
clear signatures of spontaneous scalarization when the collapse
ultimately formed a BH, but not in case of neutron-star end products,
as these were not compact enough.  Further studies using more
elaborate treatment of the microphysics and/or relaxing symmetry
assumptions are needed to determine how generic a feature this is in
core collapse scenarios.  Finally, spontaneously scalarized
configurations may also be the result of semiclassical vacuum
instabilities~\cite{Lima:2010na,Pani:2010vc,Mendes:2013ija,Landulfo:2014wra}.

The nonradial oscillation modes of spontaneously scalarized,
nonrotating stars were studied by various
authors~\cite{Sotani:2004rq,Sotani:2005qx,Sotani:2014tua,Silva:2014ora}. The
bottom line is that the oscillation frequencies can differ
significantly from their GR counterparts if spontaneous scalarization
modifies the equilibrium properties of the star (e.g., the mass-radius
relation) by appreciable amounts. However, current binary pulsar
observations yield very tight constraints on spontaneous scalarization
-- implying in particular that $\beta\gtrsim -4.5$ -- and the
oscillation modes of scalarized stars for viable theory parameters are
unlikely to differ from their GR counterparts by any measurable
amount.  Note, however, that the binary pulsar constraints on $\beta$
apply to the case of massless ST theories. For massive scalars, much
larger (negative) $\beta$ and correspondingly stronger effects on the
structure and dynamics of compact objects may be possible
\cite{Ramazanoglu:2016kul}.
\begin{figure*}[t]
\begin{center}
\begin{tabular}{ll}
\includegraphics[width=0.5\textwidth]{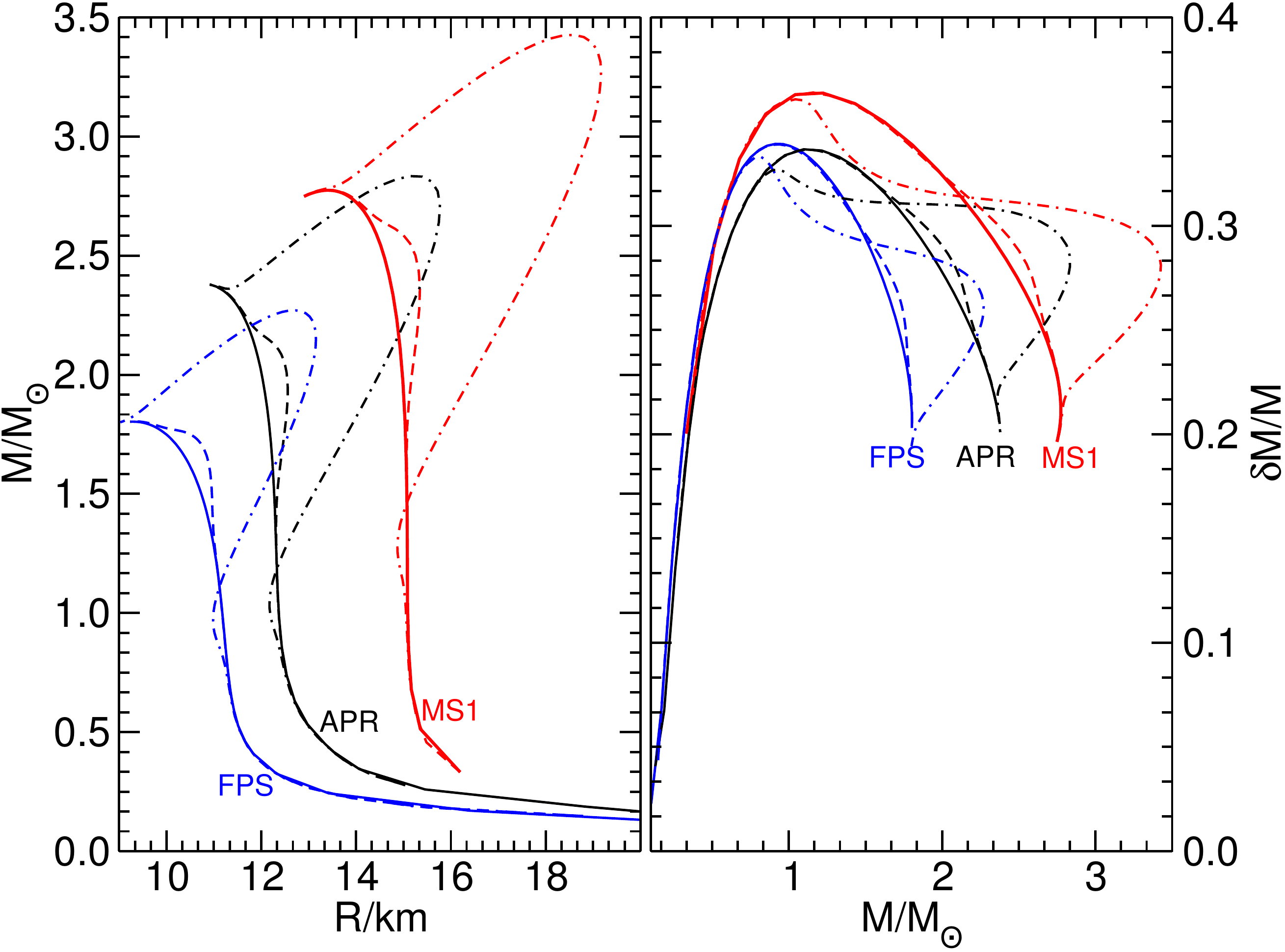}&
\includegraphics[width=0.5\textwidth]{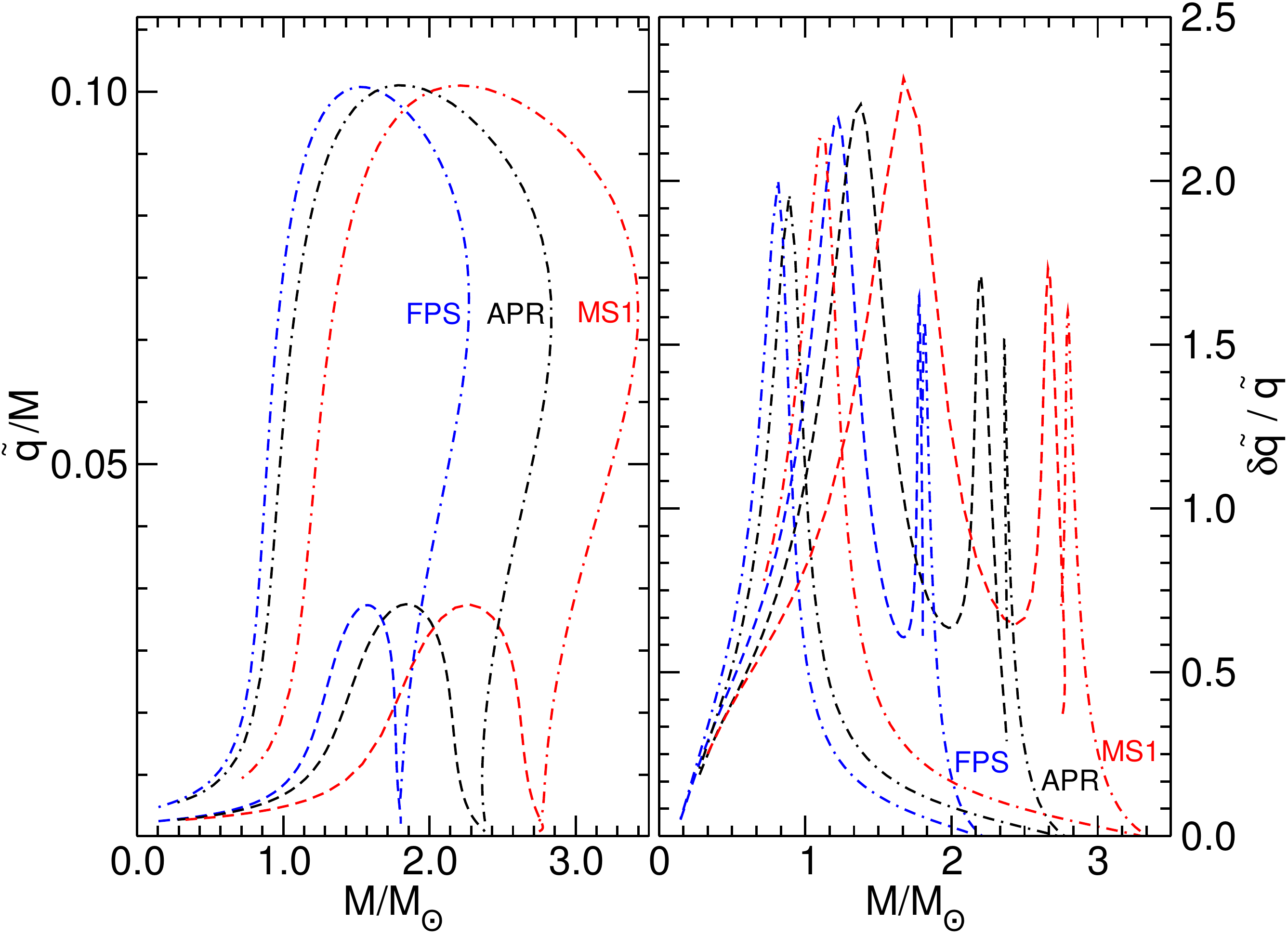}\\
\includegraphics[width=0.5\textwidth]{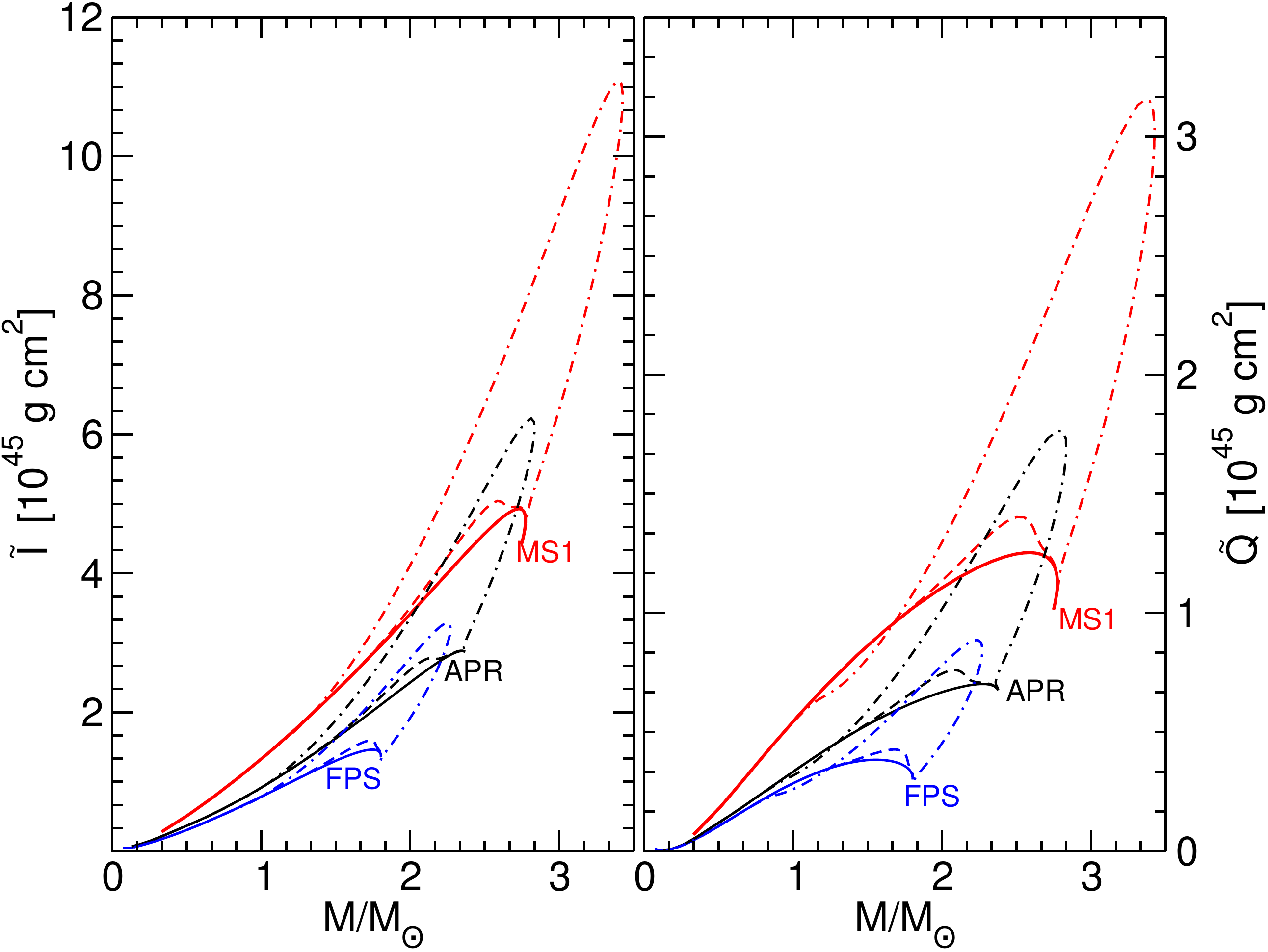}&
\includegraphics[width=0.5\textwidth]{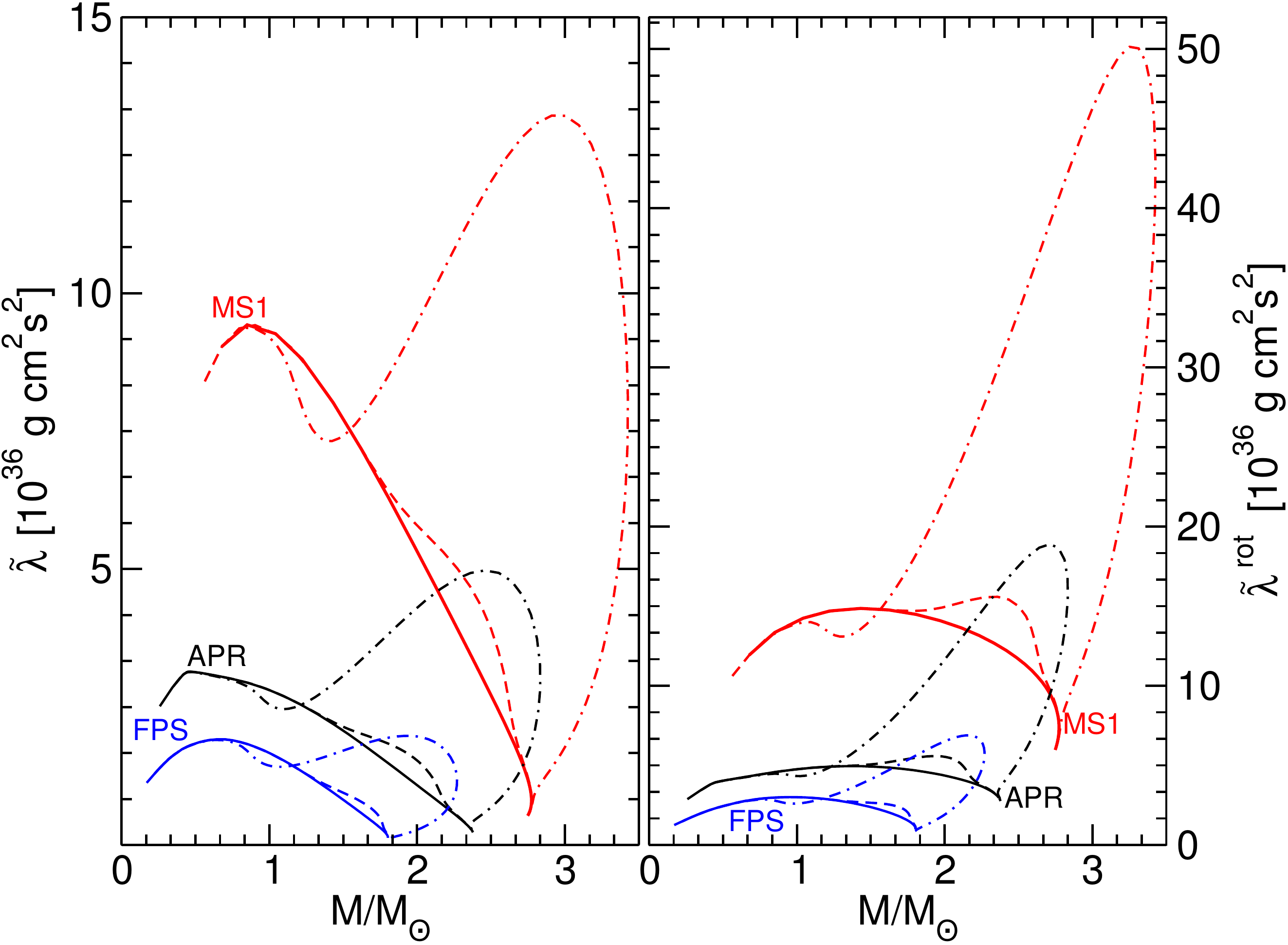}\\
\end{tabular}
\caption{NS configurations in GR (solid lines) and in two
  scalar-tensor theories defined by Eq.~\eqref{STactionE} with
  $A(\varphi)\equiv e^{\frac{1}{2}\beta_0 \varphi^2}$ and
  $V(\varphi)\equiv0$. Dashed lines refer to $\beta_0=-4.5$,
  $\varphi_0^\infty/\sqrt{4\pi}=10^{-3}$; dash-dotted lines refer to
  $\beta_0=-6$, $\varphi_0^\infty/\sqrt{4\pi}=10^{-3}$.  Each panel 
  shows results for three different EOS models (\texttt{FPS}, 
  \texttt{APR} and \texttt{MS1}).
  Top-left panel, left inset: relation between the nonrotating mass
  $M$ and the radius $R$ in the Einstein frame. Top-left panel, right inset:
  relative mass correction $\delta M/M$ induced by rotation at the
  Keplerian limit as a function of the mass $M$ of a nonspinning star
  with the same central energy density.
  Top-right panel, left inset: scalar charge $\tilde{q}/M$ as a function
  of $M$. Top-right panel, right inset: relative correction to the
  scalar charge $\delta \tilde{q}/\tilde{q}$ induced by rotation as a
  function of $M$.
  Bottom-left panel: Jordan-frame moment of inertia $\tilde{I}$ (left
  inset) and Jordan-frame quadrupole moment $\tilde{Q}$ (right inset)
  as functions of $M$.
  Bottom-right panel: Jordan-frame tidal ($\tilde{\lambda}$) and
  rotational ($\tilde{\lambda}^{\rm rot}$) Love numbers as functions
  of $M$. [From Pani and Berti\cite{Pani:2014jra}.]
  \label{fig:NS_ST2}}
\end{center}
\end{figure*}

Spinning NSs at first order in the Hartle-Thorne slow-rotation
approximation were studied by Damour and
Esposito-Far\`ese~\cite{Damour:1996ke} and later by
Sotani~\cite{Sotani:2012eb}. At first order in rotation, the scalar
field only affects the moment of inertia, mass and radius of the
NS. Second-order calculations~\cite{Pani:2014jra} are necessary to
compute corrections to the spin-induced quadrupole moment, tidal and
rotational Love numbers, as well as higher-order corrections to the NS
mass and to the scalar charge. Figure~\ref{fig:NS_ST2} shows
representative examples of the properties of NSs in a scalar-tensor
theory with spontaneous scalarization at second order in the rotation
parameter.

Rapidly rotating NSs in scalar-tensor theories were recently
constructed~\cite{Doneva:2013qva} by extending the {\tt RNS}
code~\cite{Stergioulas:2003yp}. Scalarization effects are stronger --
and deviations from GR are larger -- for rapidly spinning
NSs~\cite{Doneva:2014uma,Doneva:2014faa}.
Therefore, despite the tight binary pulsar bounds, it is still
possible that spontaneous scalarization may occur in rapidly rotating
stars.

Old, isolated NSs, as well as the NSs whose inspiral and merger we
expect to observe with GW detectors, are expected to be rotating well
below their mass-shedding limit. However these considerations may not
apply just before merger, where the rotational frequencies of each NS
may approach the mass-shedding limit. In these conditions, numerical
simulations have also recently revealed the possibility of ``dynamical
scalarization'' -- a growth of the scalar field that may significantly
affect the waveform near merger, and potentially be
detectable~\cite{Barausse:2012da,Palenzuela:2013hsa,Shibata:2013pra,Taniguchi:2014fqa,Sampson:2014qqa}.

\begin{figure*}[t]
\begin{center}
\includegraphics[width=\textwidth]{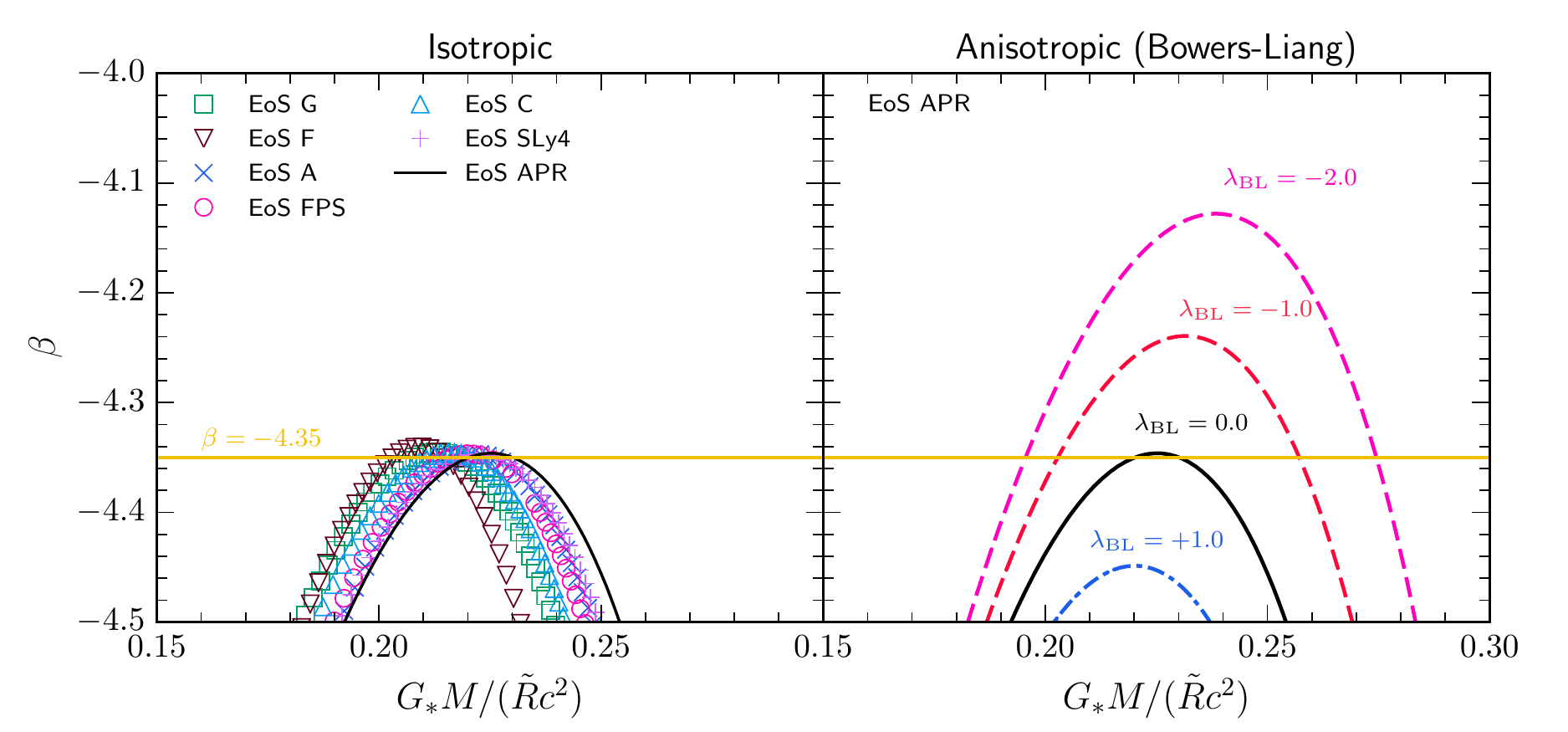}
\caption{Effect of anisotropy on the scalarization threshold. [Adapted
  from Silva et al.~\cite{Silva:2014fca}.]
  \label{fig:aniso}}
\end{center}
\end{figure*}

A more exotic mechanism to amplify the effects of scalarization is
anisotropy in the matter composing the
star~\cite{Silva:2014fca}. Nuclear matter may be anisotropic at very
high densities, where the nuclear interactions must be treated
relativistically and phase transitions (e.g. to pion condensates or to
a superfluid state) may occur. For example, Nelmes and
Piette~\cite{Nelmes:2012uf} recently considered NS structure within
the Skyrme model -- a low energy, effective field theory for quantum
chromodynamics (QCD) -- finding significant anisotropic strains for
stars with mass $M \gtrsim 1.5M_\odot$ (see also work by Adam et
al.\cite{Adam:2014dqa,Adam:2015lpa}).  The effect of anisotropy is
shown in Fig.~\ref{fig:aniso}. For illustration, in the figure we
adopt a very simple model developed in the seventies by Bowers and
Liang \cite{BowersLiang1974}, where the degree of anisotropy is
parametrized by a parameter $\lambda_{\rm BL}$.
The left panel shows the critical threshold for scalarization as a
function of stellar compactness for several ``ordinary'' (isotropic)
EOSs: the EOS has almost no effect on the critical threshold for
scalarization, which is always around $\beta=-4.35$.
The fact that scalarization is only possible when
$\beta\lesssim -4.35$ was first shown by Harada\cite{Harada:1998}
using catastrophe theory.
In the right panel, on the other hand, we show that the critical
$\beta$ for scalarization (and, as it turns out, also the effects of
scalarization on macroscopic NS properties) increases (decreases) when
the tangential pressure is bigger (smaller) than the radial pressure.

An interesting feature of the Bowers-Liang models is that it allows
for stellar configurations with compactness approaching the
Schwarzschild limit $r=2M$.  Yagi and Yunes used this observation to
study the recently found ``I-Love-Q'' universal relations -- which
relate bulk NS properties such as the moment of inertia, spin-induced quadrupole
moment and tidal deformability in an EOS-independent way -- as NSs
approach the BH limit \cite{Yagi:2015upa,Yagi:2015hda,Yagi:2016ejg}.

\subsubsection{Black hole hair?}

The phenomenology of scalar-tensor theory in vacuum spacetimes, such
as BH spacetimes, is less interesting. When the matter action $S_M$
can be neglected, the Einstein-frame formulation of the theory is
equivalent to GR minimally coupled to a scalar field. BHs in
Bergmann-Wagoner theories satisfy the same no-hair theorem as in GR,
and thus the stationary BH solutions in the two theories
coincide~\cite{Hawking:1972qk,Heusler:1995qj,Sotiriou:2011dz}. Moreover,
dynamical (vacuum) BH spacetimes satisfy a similar generalized no-hair
theorem: the dynamics of a BH binary system in Bergmann-Wagoner theory
with vanishing potential are the same as in GR~\cite{Damour:1992we},
up to at least $2.5$ PN order for generic mass
ratios~\cite{Mirshekari:2013vb} and at any PN order in the extreme
mass-ratio limit~\cite{Yunes:2011aa}.

If there is more than one massive real scalar field, however, or at
least one massive complex scalar field, the situation concerning
stationary BH solutions can be very different: axisymmetric, hairy BHs
\textit{do
  exist}\cite{Herdeiro:2014goa,Herdeiro:2015gia,Herdeiro:2015waa}, as
will be reviewed in
Section~\ref{sec_superradiance}. Tensor-multi-scalar theories have
indeed received more attention in the recent literature, as we now
discuss.

\subsection{Tensor-multi-scalar theories}
\label{sec:TMS}

A natural generalization of the Bergmann-Wagoner formulation~\eqref{STactionJ} consists in including more than one
scalar field coupled with gravity. The action of tensor-multi-scalar (TMS)
gravity~\cite{Damour:1992we,Horbatsch:2015bua} is:
\begin{align}
S\,=\,\,& \frac1{16\pi}\int d^4x\sqrt{-g}\left(
F(\phi)R-\gamma_{ab}(\phi)g^{\mu\nu}\partial_\mu\phi^a
\partial_\nu\phi^b-V(\phi)\right)+S_M[\Psi,\,g_{\mu\nu}]\,,
\label{STactionJ_multi}
\end{align}
where $F$ and $V$ are functions of the $N$ scalar fields $\phi^a$
($a=1\ldots N$). The scalar fields live on a manifold (the {\it target
  space}) with metric $\gamma_{ab}(\phi)$. The
action~\eqref{STactionJ_multi} is invariant not only under spacetime
diffeomorphisms, but also under target-space diffeomorphisms,
i.e.~scalar field redefinitions. TMS theories are more complex than
theories with a single scalar field, since the geometry of the target
space can affect the dynamics.

The simplest extension of a ST theory with a single real scalar field
is a theory with two real scalar fields.  If we work, equivalently,
with a single complex scalar $\varphi$ instead, the action reduces to
\begin{align}
S = \frac{1}{4 \pi G_{\star}} \int d^{4}x \sqrt{-g} \left[ \frac{R}{4}
-g^{\mu \nu} \gamma(\varphi,\bar{\varphi}) \nabla_{\mu}\bar{\varphi}
\, \nabla_{\nu}\varphi - V(\varphi,\bar{\varphi}) \right]
+ S_{\rm m}[A^{2}(\varphi,\bar{\varphi})g_{\mu \nu} ; \Psi] \,, \nonumber \\
\label{eq:action_ef_complex_single}
\end{align}
%
Hereafter we assume that the potential vanishes,
i.e. $V(\varphi,\bar{\varphi})=0$, and that the target space is
maximally symmetric. Upon stereographic projection and field
redefinition\cite{Horbatsch:2015bua} the target-space metric can be
written as
\begin{equation}
  \gamma(\varphi,\bar{\varphi}) = \frac{1}{2} \left( 1 + \frac{\bar{\varphi}\varphi}{4\gothr^2}\right)^{-2} \,,
  \label{targmetr}
\end{equation}
where $\gothr$ is the radius of curvature of the target-space
geometry.  For a spherical geometry we have $\gothr^2>0$, for a
hyperbolic geometry $\gothr^2<0$, and in the limit
$\gothr \rightarrow \infty$ the geometry is flat.

The function $A(\varphi,\bar{\varphi})$ determines the scalar-matter
coupling. What enters the field equations is actually the function
$\kappa$, defined as
\begin{equation}
\kappa(\varphi,\bar{\varphi}) \equiv 2 \left( 1 + \frac{\bar{\varphi} \varphi}{4 \radcurv^{2}}\right)
\frac{\partial \log A(\varphi,\bar{\varphi})}{\partial \varphi}.
\label{eq:kappa_def}
\end{equation}
Without loss of generality we assume that far away from the
source the field vanishes: $\varphi_\infty=0$.  We then expand the
function $\log A$ in a series about $\varphi=0$:
\begin{equation}
\log A(\varphi, \bar{\varphi}) =\alpha^\ast\varphi+\bar{\alpha}^\ast\bar{\varphi}+
\frac{1}{2}\beta_0 \varphi \bar{\varphi} +
\frac{1}{4} \beta^\ast_1 \varphi^2  +
\frac{1}{4}\bar{\beta}^\ast_1 \bar{\varphi}^2 +\dots\,,
\label{eq:A_expansion}
\end{equation}
where $\beta_0$ is real, while $\alpha^\ast$ and $\beta^\ast_1$ are in
general complex numbers.  Redefine
$\beta^\ast_1 \equiv \beta_1e^{\ii \theta}$, where $\theta$ is chosen
such that $\beta_1$ is real.  Then, after defining
$\alpha^\ast \equiv \alpha e^{\ii \theta/2}$ and a new field
$\psi \equiv \varphi e^{\ii \theta/2}$, the field equations become
%
\begin{align}
R_{\mu\nu} &= 2\left( 1 + \frac{\bar{\psi}\psi}{4\gothr^2} \right)^{-2}\partial_{(\mu}\bar{\psi} \partial_{\nu)}\psi +
8\pi G_{\star}\left( T_{\mu\nu} - \frac{1}{2}Tg_{\mu\nu} \right) \,,
\label{Einstein_psi} \\
\Box\psi &= \left( \frac{2\bar{\psi}}{\bar{\psi}\psi+4\gothr^2}\right)g^{\mu\nu} \partial_{\mu}\psi\partial_{\nu}\psi
- 4\pi G_{\star}\left( 1+ \frac{\bar{\psi}\psi}{4\gothr^2}\right)
\bar{\kappa}(\psi,\bar{\psi})\,T\,,
\label{eq:psi}
\end{align}
%
where 
\bea
\log A(\psi, \bar{\psi}) &=&
\alpha\psi+\bar{\alpha}\bar{\psi}+
\frac{1}{2}\beta_0 \psi \bar{\psi} +
\frac{1}{4}\beta_1 \psi^2  +
\frac{1}{4}\beta_1 \bar{\psi}^2
+\dots \,.
\nonumber\\
&=&
\alpha\psi+\bar{\alpha}\bar{\psi}+
\frac{1}{2} \left[
(\beta_0+\beta_1) {\rm Re}[\psi]^2 +
(\beta_0-\beta_1) {\rm Im}[\psi]^2  \right]\,,   \label{cf:realsplit}
\label{eq:logA}
\eea
and in the second line we have split the field $\psi$ into real and
imaginary parts: $\psi\equiv {\rm Re}[\psi] + \ii \,{\rm Im}[\psi]$.
The structure of this theory when $\alpha=0$ is determined by three
real parameters: $\beta_0+\beta_1$, $\beta_0-\beta_1$ and the
target-space curvature $\gothr^2$. When $\alpha\neq0$, two further
parameters ($|\alpha|$ and $\arg\alpha$) are necessary to define the
theory.

This two-scalar model is the simplest generalization of the
spontaneous scalarization model by Damour and
Esposito-Far\`ese~\cite{Damour:1993hw}.
Note that the quantity
$|\alpha|^2 \equiv \alpha\bar{\alpha}\equiv
\rm{Re}[\alpha]^2+\rm{Im}[\alpha]^2$
is strongly constrained by observations, similarly to the
single-scalar case. However, in TMS theories $\alpha$ is a complex
quantity and its argument, $\arg\alpha$, is unconstrained in the
weak-field regime. When $\alpha=0$, the conformal coupling at second
order in $\psi$ reduces to
\begin{equation}
\log A(\psi, \bar{\psi}) =
\frac{1}{2}\beta_0 \psi \bar{\psi} +
\frac{1}{4}\beta_1 \psi^2  +
\frac{1}{4}\beta_1 \bar{\psi}^2 \,.
\label{cf0}
\end{equation}

\begin{figure}
  \includegraphics[height=300pt,clip=true]{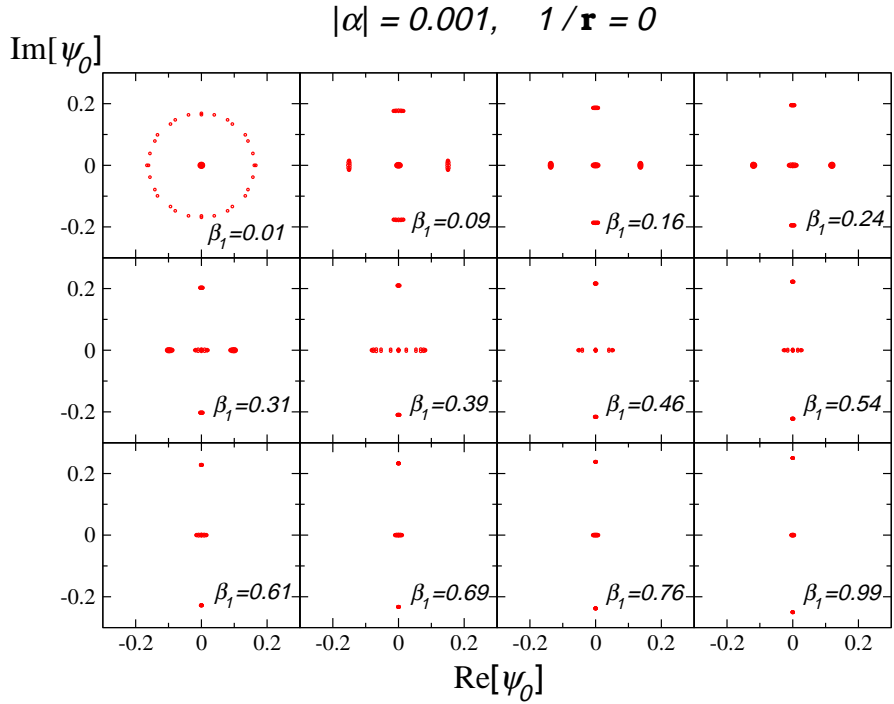}
  \caption{ {\it Scalar field amplitudes in the full TMS theory.}
    Real and imaginary part of the scalar field amplitude at the
    stellar center $\psi_0$ for stellar models with $\beta_0=-5$,
    $|\alpha|=0.001$ and fixed baryon mass $M_B = 1.8~M_{\odot}$. The
    different panels show the solutions found for different values of
    $\beta_1$, as indicated in each panel. In each case, we vary the
    phase of $\alpha$ from $0$ to $2\pi$ in steps of $\pi/6$.}
  \label{fig:bm5_alpha0001_iR0}
\end{figure}

Compact stars in theories with $\alpha=0$ and $\alpha\neq0$ are rather
different. When $\alpha = 0$ and $\beta_{1} \neq 0$, the theory is
invariant under the symmetries $\psi \to \bar{\psi}$ and
$\psi \to - \psi$. 
In this case, we only found
solutions where either the real or the imaginary part of the scalar
field has a non-trivial profile; these theories are effectively
equivalent to single-scalar theories.

When $\alpha\neq0$ the situation is more interesting, as shown in
Fig.~\ref{fig:bm5_alpha0001_iR0}. Introduction of
$\alpha \in \mathbb{R}$ partially breaks this symmetry down to
conjugation only, whereas introduction of $\alpha \in \mathbb{C}$
fully breaks the symmetry. Now GR configurations are not solutions of
the field equations. In particular, a constant (or zero) scalar field
does not satisfy Eq.~\eqref{eq:psi} when $T\neq 0$.  Therefore it is
not surprising that when $\alpha\neq0$ we can find solutions with two
nontrivial scalar profiles even when $\beta_0=\beta_1=0$. A more
interesting question is whether there are stellar configurations in
which both scalar fields have a large amplitude. These
``biscalarized'' solutions are absent in the $\alpha=0$ case, but as
it turns out they exist when $\alpha \neq 0$. For concreteness, in the
figure we set $|\alpha|=10^{-3}$, so that the theory is compatible
with experimental bounds from binary pulsar observations. We set
$1/\gothr=0$ (i.e., for simplicity we consider a flat target space),
we fix $\beta_0=-5$, and we vary $\arg\alpha$ in the range
$[0,\,2\pi]$ in steps of $\pi/6$.  As shown in
Fig.~\ref{fig:bm5_alpha0001_iR0} -- where dots denote the real and
imaginary parts of the central value of the scalar field $\psi_0$ for
which solutions were found -- there are several solutions where {\em
  both} the real and imaginary part of the scalar field are
nonzero. The solutions are at least approximately $O(2)$ symmetric
when $\beta_1\sim |\alpha|$. The symmetry is broken (the solution
``circles'' turn into ``crosses'') when $\beta_1\gg |\alpha|$, and the
cross-like shape of the scalarized solutions in the
$({\rm Re}[\psi_0]$, ${\rm Im}[\psi_0])$ plane collapses towards a set
of solutions on the vertical line ${\rm Re}[\psi_0]=0$ for the larger
values of $\beta_1$ (bottom panels in
Fig.~\ref{fig:bm5_alpha0001_iR0}).  For the case $\alpha=0$, it is
easy to see from Eq.~\eqref{cf:realsplit} that spontaneous
scalarization of ${\rm Re}[\psi]$ occurs (in analogy with the
single-field case) if $\beta_0 + \beta_1 \lesssim -4.35$, and that
scalarized models with a large imaginary part ${\rm Im}[\psi]$ exist
if $\beta_0 - \beta_1 \lesssim -4.35$. Our biscalarized models have
been calculated for fixed $\beta_0=-5$. For $\beta_1 \gtrsim 0.65$ we
therefore enter the regime where $\beta_0+\beta_1 \gtrsim -4.35$, and
we no longer expect to find models with strongly scalarized
${\rm Re}[\psi]$.  The condition $\beta_0-\beta_1 \lesssim -4.35$ for
scalarization of ${\rm Im}[\psi]$, however, remains satisfied, so that
scalarized models should cluster close to the
${\rm Re}[\psi_0]$--axis. This is indeed observed in the bottom panels
of Fig.~\ref{fig:bm5_alpha0001_iR0}.  A more detailed investigation of
the phenomenology of these models is underway.

\subsection{Horndeski theories}
\label{sec:Horndeski}

Besides the obvious addition of one or more scalar field(s), a second
possibility to generalize scalar-tensor theories of the
Bergmann-Wagoner type has recently attracted a great deal of
attention. The theory in question was first formulated by
Horndeski~\cite{Horndeski:1974wa}, and it is the most general
single-scalar theory with second-order field equations.  In ``modern''
notation, the action of Horndeski gravity can be written in terms of
Galileon interactions~\cite{Deffayet:2011gz} as
\begin{equation}
S=\frac{1}{8\pi}\sum_{i=2}^{5}\int d^{4}x\sqrt{-g}{\cal L}_i\ ,
\label{eq:action}
\end{equation}
where
\begin{subequations}
\begin{align}
{\cal L}_2&=G_2\ ,\\
{\cal L}_3&=-G_{3}\square\phi\ ,\\
{\cal L}_4&=G_{4}R+G_{4\tn{X}}\left[(\square\phi)^2-\phi_{\mu\nu}^2\right]\ ,\\
{\cal L}_5&=G_{5}G_{\mu\nu}\phi^{\mu\nu}-\frac{G_{5\tn{X}}}{6}\left[(\square\phi)^3+2\phi_{\mu\nu}^3 -3\phi_{\mu\nu}^2\square\phi\right]\ .
\label{eq:lagrangean}
\end{align}
\end{subequations}
The functions $G_{i}=G_{i}(\phi,X)$ depend only on the scalar
field $\phi$ and its kinetic energy, $X=-\partial_\mu\phi\partial^\mu\phi/2$.
For brevity we have also defined the shorthand notation
$\phi_{\mu\dots\nu}\equiv \nabla_\mu\dots\nabla_\nu\phi$,
$\phi_{\mu\nu}^2 \equiv \phi_{\mu\nu}\phi^{\mu\nu}$,
$\phi_{\mu\nu}^3 \equiv
\phi_{\mu\nu}\phi^{\nu\alpha}\phi^{\mu}{_{\alpha}}$
and $\Box\phi\equiv g^{\mu\nu} \phi_{\mu\nu}$.

An attractive feature of Horndeski gravity is its generality. The
theory includes a broad spectrum of phenomenological dark energy
models, as well as modified gravity theories with a single scalar
degree of freedom. Some important special limits of the theory are
listed below:

\begin{enumerate}
\item GR corresponds to choosing $G_4=1/2$ and $G_2=G_3=G_5=0$.
\item When $G_4=F(\phi)$ and all other $G_{i}$'s are zero we recover a
  scalar-tensor theory with nonminimal coupling of the form
  $F(\phi)R$. Therefore Brans-Dicke theory and $f(R)$ gravity are
  special cases of Horndeski gravity.
\item A theory that we will consider in some detail below, namely
  Einstein-dilaton-Gauss-Bonnet (EdGB) gravity, has the action
  \begin{equation}\label{eq:EdGBaction}
    S=\frac{1}{16\pi}\int d^4x
\sqrt{-g}\left(R+X-V(\phi)+\xi(\phi) R^2_{\tn{GB}}\right)\ ,
  \end{equation}
  where $V(\phi)$ is the scalar potential, $\xi(\phi)$ is a coupling
  function and
\begin{equation}\label{eq:RGB}
  R^2_\tn{GB}=R^2-4R_{\mu\nu}R^{\mu\nu}+R_{\alpha\beta\gamma\delta}R^{\alpha\beta\gamma\delta}
\end{equation}
is the Gauss-Bonnet invariant. This theory can be recovered with the
choices\cite{Kobayashi:2011nu,Maselli:2015yva}
\begin{subequations}
\begin{align}
G_2&=\frac{X}{2}-\frac{V}{2}+4\xi^{(4)} X^2(3-\ln X)\,,\quad
G_3=2\xi^{(3)}X(7-3\ln X)\,,\label{EDGBK}\\
G_4&=\frac{1}{2}+2\xi^{(2)}X(2-\ln X)\,, \quad
G_5=-2\xi^{(1)}\ln X\,,\label{EDGBG5}
\end{align}
\end{subequations}
where we have defined
$\xi^{(n)}\equiv \partial^n \xi/\partial
\phi^n$~\cite{Kobayashi:2011nu}.
\item A theory with nonminimal derivative coupling between the scalar
  field $\phi$ and the Einstein tensor $G_{\mu\nu}$ (the ``John''
  Lagrangian in the language of the so-called ``Fab Four''
  model~\cite{Charmousis:2011bf,Charmousis:2011ea}), with action
\begin{equation}\label{eq:nonminact}
S=\int d^{4}
x\sqrt{-g}\left[\zeta R+2\beta X+\eta G^{\mu\nu}\phi_\mu\phi_\nu
-2\Lambda_0\right]\,,
\end{equation}
can be constructed by setting
\begin{equation}
G_{2}=-2\Lambda_0+2\beta X\ , \quad
G_4=\zeta+\eta X\ , \quad
G_3=G_5=0\ ,
\end{equation}
where $\Lambda_0$, $\eta$, $\zeta$ and $\beta$ are
constants. Incidentally, a coupling of the form
$G^{\mu\nu}\phi_{\mu}\phi_{\nu}$ can also be obtained by setting
$G_5=-\phi$ and integrating by parts~\cite{Kobayashi:2014eva}
\label{itm:phieins}
\item The Lagrangian ${\cal L}_2$ corresponds to the k-essence
  field~\cite{ArmendarizPicon:2000ah,ArmendarizPicon:1999rj,Alishahiha:2004eh}.
  For this reason, some of the literature denotes the function $G_2$
  by the letter $K$.
\label{itm:kess}
\item\label{itm:covgal} The covariant Galileon
  model~\cite{Deffayet:2009wt} corresponds to setting $G_2=-c_2X$,
  $G_3=-c_3 X/M^3$, $G_4=M_{\rm Pl}^2/2-c_4X^2/M^6$ and
  $G_5=3c_5X^2/M^9$, where the $c_i$ ($i=2,\ldots,5$) are constants
  and $M$ is a constant with dimensions of mass.
\end{enumerate}

Because of the generality of Horndeski gravity, a comprehensive review
of compact objects would inevitably have to discuss important
subclasses that have been studied for a long
time~\cite{Berti:2015itd}. A more specific review of compact objects
in the subclasses \ref{itm:phieins}--\ref{itm:covgal} can be found in
this same volume\cite{Silva:2016smx}, and some examples are also
discussed in another review\cite{Herdeiro:2015waa}.  In the next
paragraph we complement these reviews focusing on recent work in EdGB
gravity.

\subsection{Einstein-dilaton-Gauss-Bonnet gravity}
\label{sec:EdGB}

In EdGB gravity~\cite{Kanti:1995vq} (see Section~\ref{sec:Horndeski}),
the Gauss-Bonnet invariant (\ref{eq:RGB}) is coupled with a scalar
field ~\footnote{The normalization of the scalar is different from
  those in Sections~\ref{sec:BW} and~\ref{sec:TMS}, by a factor $2$.}.
The resulting action of EdGB gravity, Eq.~(\ref{eq:EdGBaction}),
is a special case of Horndeski gravity, as discussed in
Section~\ref{sec:Horndeski}. With the choice $\xi(\phi)=\alpha\phi$
this theory is shift-symmetric, and it has been
shown~\cite{Sotiriou:2013qea} that it is the only shift-symmetric
Horndeski gravity theory in which the no-hair theorems do not hold.

EdGB gravity can also be seen as belonging to a different class of
modified gravity: that of quadratic gravity
theories~\cite{Yunes:2011we,Pani:2011gy}, in which quadratic curvature
terms are included in the action. EdGB gravity holds a special place
as the only quadratic gravity theory with equations of motion of
second differential order. Other theories of quadratic curvature
gravity (e.g. dynamical Chern-Simons
gravity~\cite{jackiw:2003:cmo,Alexander:2009tp}) have equations of
motion of higher differential order, and are then subject to
Ostrogradski's instability~\cite{Woodard:2006nt}. In order to avoid
this instability they should be considered as effective theories,
obtained as truncations of more general theories. In other words, EdGB
gravity is consistent for any value of the coupling constant, while
other quadratic gravity theories should only be considered in the
weak-coupling limit. Note also that the EdGB term without the coupling
to a scalar field would be trivial, since $R^2_{\tn{GB}}$ is a total
derivative.

Including a quadratic curvature term in the action is an interesting
modification of GR, for a variety of reasons. First of all, this is
the simplest way to modify the strong-curvature regime of gravity,
and, second, it is also a way to circumvent no-hair theorems (see the
discussion in Sections~\ref{sec:BW} and~\ref{sec_superradiance} for
different ways to grow BH hair). Moreover, quadratic curvature terms
can make the theory renormalizable~\cite{Stelle:1976gc}. In
particular, the EdGB term naturally arises in low-energy effective
string theories~\cite{Gross:1986mw,Moura:2006pz} when $\xi(\phi)=\alpha e^{\phi}/4$.

Hereafter we consider EdGB gravity with $\xi(\phi)=\alpha e^{\phi}/4$.
The first BH solution of this theory, derived about 20 years ago by
Kanti et al.~\cite{Kanti:1995vq}, is a numerical solution describing a
spherically symmetric BH. The solution has scalar hair, i.e. a
non-trivial configuration of the scalar field, but only {\it secondary
  hair} (the scalar charge $D$ is determined by the mass $M$, and
hence is not a free parameter.  It can be shown that Kanti's solution
only exists for~\cite{Kanti:1995vq,Pani:2009wy}
\begin{equation}
0<\alpha/M^2\lesssim0.691\label{boundalpha}\,,
\end{equation}
where $M$ is the BH mass. The best observational bound on the coupling
parameter is $\alpha\lesssim47\,M_\odot^2$~\cite{Yagi:2012gp}. This
bound is weaker than the theoretical bound~\eqref{boundalpha} for BHs
with $M\lesssim8.2\,M_\odot$~\cite{Maselli:2014fca}.

In recent years, numerical solutions describing slowly
rotating~\cite{Pani:2009wy} and rapidly
rotating~\cite{Kleihaus:2011tg} BHs have been derived. These solutions
describe stationary BHs for all values of the mass and the spin, and
for all values of the coupling parameter in the allowed
range~\eqref{boundalpha}.  However, these solutions require a
numerical integration for each set of parameters.  In order to devise
and implement observational tests based on astrophysical or GW
observations (for instance, for Monte Carlo data analysis), an
approximate, analytical solution can be more useful than numerical
solutions.

Analytical BH solutions in EdGB theory have been determined as
perturbative expansions in the dimensionless coupling parameter
$\zeta=\alpha/M^2$ and the dimensionless spin ${\bar a}=J/M^2$, at
order $O(\zeta^2,{\bar a}^0)$~\cite{Mignemi:1992nt,Yunes:2011we},
$O(\zeta^2,{\bar a}^1)$~\cite{Pani:2011gy},
$O(\zeta^2,{\bar a}^2)$~\cite{Ayzenberg:2013wua}, and finally at order
$O(\zeta^7,{\bar a}^5)$~\cite{Maselli:2015tta}. For a slowly rotating
BH, the solution derived in~\cite{Maselli:2015tta} reproduces the most
relevant geodesic quantities (the ISCO location and the epyciclic
frequencies) within $1\%$, for the entire allowed
range~\eqref{boundalpha} of the coupling parameter.

Astrophysical observations from the near-horizon region of BHs can
allow tests of GR against modified theories (such as EdGB gravity)
which predict deviations in the strong-field, high-curvature
regime. Indeed, near the horizon of stellar-mass BHs the spacetime
curvature is very large, and (for sufficiently large values of
$\alpha$) BH solutions in EdGB theory may be significantly different
from the Kerr solution. These deviations can affect observable
quantities, such as the quasi-periodic oscillations (QPOs) observed in
the $X$-ray flux of accreting BHs~\cite{Maselli:2014fca}. Indeed, in
many astrophysical models the frequencies of these QPOs are
appropriate combinations of the epicyclic frequencies of the
(near-horizon) BH geodesics, in which the strong-field regime of
gravity is manifest.  Therefore future large-area $X$-ray telescopes
such as LOFT~\cite{feroci:2012qh} could set constraints on the
coupling parameter $\alpha$. For instance, the detection of two QPO
triplets from a BH with $M=5.3\,M_\odot$ and $\bar a=0.2$ by a
detector having the LOFT design sensitivity could exclude the range
$\zeta\gtrsim0.4$ with $3\sigma$ confidence~\cite{Maselli:2014fca}.

\section{Implications of superradiant instabilities for fundamental physics and astrophysics}
\label{sec_superradiance}
In the previous section we focused on specific modifications of
Einstein's gravity and on the different physical consequences, as well
as compact object solutions, that arise in these models. Perhaps
somewhat more surprisingly, even {\em within} Einstein's gravity,
considering simple fundamental fields that satisfy the energy
conditions can also lead to new types of compact objects, with
interesting physical consequences. In this section we will review
these recent developments, that are related to the complex phenomenon
known as \textit{superradiance}.\cite{Brito:2015oca}

\subsection{Setup}
Einstein's GR minimally coupled to fundamental fields, such as massive
scalars or vectors, is described by the Lagrangian
\be
{\cal L}=\frac{R}{\kappa} - \frac{F^*_{\mu\nu}F^{\mu\nu}}{4}- \frac{\mu_V^2}{2}A^*_{\nu}A^{\nu}-\frac{g^{\mu\nu}}{2}\phi^{\ast}_{,\mu}\phi^{}_{,\nu} - \frac{\mu_S^2}{2}\phi^{\ast}\phi\, \ .\label{eq:MFaction}
\ee
We have defined $\kappa=16\pi$, and
$F_{\mu\nu} \equiv \nabla_{\mu}A_{\nu} - \nabla_{\nu} A_{\mu}$ is the
Maxwell tensor. Both the scalar and vector fields are assumed to be
complex, for reasons that will become clear soon. The mass $m_B$ of
the bosons under consideration is related to the mass parameters above
through $\mu_{S,V}=m_{B}/\hbar$. By ``fundamental'' we mean fields
which are not effective descriptions of other microscopic degrees of
freedom. For most of the analysis below, however, the true nature of
these fields (i.e., whether they are truly fundamental or rather a
coarse-grained representation of more fundamental degrees of freedom),
is not relevant. Each of them is completely equivalent to two real
scalar or vector fields, but some of our considerations below apply
equally well to one or many {\em real} scalar and vector fields.

The theories represented by this action are relevant for several
reasons. Because they are simple, they can be thought of as proxies
for more complex interactions, of which they would be faithful models
in certain regimes (presumably when higher-order interactions are
negligible). Fundamental bosons also play a key role in particle
physics. For instance they could describe the axion or axion-like
particles, originally intended to solve the strong-CP problem in QCD,
which recently gained prominence as dark-matter
candidates~\cite{Marsh:2015wka,2014NatPh..10..496S,2014PhRvL.113z1302S}. In
this context, self-gravitating solutions of fundamental fields allow
us to understand and study quantitatively the growth of dark matter
structures and their clustering inside compact
stars~\cite{Brito:2015yga,Brito:2015yfh}.

Whether or not they form a significant component of dark matter,
minimally coupled fundamental fields should obey the equivalence
principle and freely fall in the same way as standard model
fields. Thus, the most promising channel to look for their imprints
consists of gravitational interactions.

\subsection{Superradiance and superradiant instabilities}
Fundamental fields in the presence of gravity display of course a
panoply of interesting effects, such as the critical phenomena
identified in Choptuik's seminal study\cite{Choptuik:1992jv}.  In
strong gravitational fields, one of the most peculiar is
superradiance, i.e., the amplification of low-frequency waves
scattering off rotating
BHs~\cite{zeldovich1,zeldovich2,Brito:2015oca}. Superradiance is
required by the second law of thermodynamics, and is akin to tidal
acceleration in planetary
dynamics~\cite{Cardoso:2012zn}. Superradiance is active for
low-frequency, bosonic fields satisfying the {\it superradiance
  condition}
\be
\label{supercondition}
\omega < m\Omega\,,
\ee
with $m$ an integer azimuthal number and $\Omega$ the angular
frequency of the BH. The amplitude of the superradiant amplification
of any incident wave depends on the rotation $\Omega$, on the wave
frequency $\omega$ and on the field being
scattered~\cite{Teukolsky:1974yv,Brito:2015oca}.

Superradiant mechanisms can trigger instabilities in spacetimes that
are able to confine the fluctuations. In such cases, the wave is
forced to bounce back and forth, being repeatedly amplified upon
interaction with the BH, and leading to exponential growth of
linearized fluctuations. This mechanism is called a {\it black hole
  bomb}~\cite{Press:1972zz,Cardoso:2004nk,Rosa:2009ei,Hod:2013fvl},
and leads to instabilities in truly confined spacetimes like anti-de
Sitter~\cite{Cardoso:2004hs,Cardoso:2006wa,Cardoso:2013pza,Brito:2014nja,Bosch:2016vcp}.

It is interesting that the same mechanism also makes {\it Kerr BHs}
unstable under massive, scalar-field
fluctuations~\cite{Damour:1976kh,Detweiler:1980uk,Cardoso:2005vk,Dolan:2012yt},
vector-field fluctuations~\cite{Pani:2012vp,Pani:2012bp,Witek:2012tr}
or even tensor-field perturbations~\cite{Brito:2013wya}. Physically,
massive states prevent full leakage to infinity and act as an
effective barrier for low-frequency waves.

\subsection{Hairy black holes bifurcating from the Kerr solution}
Since Kerr BHs are unstable against sufficiently low frequency modes
of a massive bosonic field, a relevant question is: what is the
endpoint of the instability? While this is still an open question (but
see Section~\ref{hairyform} below), a relevant observation is the
existence of stationary, asymptotically flat BH solutions of the
model~\eqref{eq:MFaction}, which are regular on and outside the event
horizon and for which the horizon is in equilibrium with a non-trivial
scalar or vector field condensate. Moreover these BHs are continuously
connected with the Kerr solution, and as such they have been
dubbed~\textit{Kerr BHs with
  scalar~\cite{Herdeiro:2014goa,Herdeiro:2014ima,Herdeiro:2015gia} or
  Proca hair\cite{Herdeiro:2016tmi}}. They are manifestly related to
the phenomenon of superradiance, as they exist at the threshold of the
inequality~\eqref{supercondition}, and they are likely to play a role
either as endpoints or as long-lived intermediate states in the
development of the superradiant instability of Kerr BHs in the
presence of massive scalar or vector fields.

The existence of these hairy BH solutions raises three immediate questions:

{\bf (1)} ``How is it possible that stationary, asymptotically flat BH
solutions different from Kerr exist in the very simple
model~\eqref{eq:MFaction}, in view of the well-known no-hair theorems
that apply to this model (in particular the pioneering theorems due to
Bekenstein for the scalar\cite{Bekenstein:1972ny} and
Proca\cite{Bekenstein:1971hc,Bekenstein:1972ky} cases)?" (see also
Ref.~\refcite{Herdeiro:2015waa} for a review of no-hair theorems
applying to the scalar case).

{\bf (2)} ``If these hairy BH solutions are continuously connected to
the Kerr solution, then there must be an imprint of their existence
when we consider the corresponding matter fields on the Kerr
background as test fields. Is it so?"

{\bf (3)} ``Do these BHs trivialize in the limit of vanishing horizon
or is there some residual gravitating configuration in this limit?"

We shall tackle each of these questions in the following three subsections.

\subsubsection{Circumventing no-hair theorems}
The answer to ${\bf (1)}$ is simple and enlightening: theorems have
assumptions and assumptions can be dropped. In the present case, an
assumption in many of the no-hair theorems, including those of
Bekenstein, is that the metric and the matter field share the same
symmetries. This is not necessary: the spacetime and the {\em
  energy-momentum tensor} should share the same symmetries, but not
the matter field itself. This apparently innocuous observation allows
us to circumvent the simplest Bekenstein-type no-hair theorems, but
observe that it is a \textit{necessary but not sufficient ingredient.}
The reason will become clear in the following.

The metric ansatz that has been successfully used for finding
(non-extremal) Kerr BHs with scalar\cite{Herdeiro:2014goa} and
Proca\cite{Herdeiro:2016tmi} hair reads:
\begin{equation}
ds^2=-e^{2F_0(r,\theta)}N(r)dt^2+e^{2F_1(r,\theta)}\left[\frac{dr^2}{N(r)}+r^2d\theta^2\right]+r^2\sin^2\theta e^{2F_2(r,\theta)}[d\varphi-W(r,\theta)dt]^2
\label{ansatzkbhssh}
\end{equation}
where $N(r)\equiv 1-r_H/r$ and $r_H$ is a constant.  The metric is
completely determined by four functions of the spheroidal coordinates
$(r,\theta)$. These coordinates reduce to \textit{prolate} spheroidal
coordinates (rather than the more familiar \textit{oblate} spheroidal
ones, obtained in the flat-spacetime limit of the Boyer-Lindquist form
of the Kerr metric) in an appropriate Minkowski
limit.\cite{Herdeiro:2016tmi} A simple analysis shows that
$r_H=$constant surfaces are null (assuming they are regular). On these
surfaces, null orbits with $\theta=$ constant have an angular
velocity, as measured by the observer at infinity,
$\Omega_H\equiv W(r_H,\theta)$.  From the numerical solutions, it
turns out that $\Omega_H$ is $\theta$-independent and $r=r_H$ is a
Killing horizon of the Killing vector field
$\xi=\partial_t+\Omega_H\partial_\varphi$. Thus $r=r_H$ is the event
horizon. Observe that the line element~\eqref{ansatzkbhssh} admits two
independent Killing vector fields: ${\bf k}\equiv\partial_t$ and
${\bf m}=\partial_\varphi$.

On the other hand, the ``matter" ansatz that has been used to find the hairy BHs is 
\begin{equation}
\Psi(t,r,\theta,\varphi)=e^{-i\omega t+im\varphi}\phi(r,\theta)
\label{ansatzscalar}
\end{equation}
for the scalar case\cite{Herdeiro:2014goa}, and
\begin{equation}
A(t,r,\theta,\varphi)=e^{-i\omega t+im\varphi}\{i[V(r,\theta)dt+H_3(r,\theta)\sin\theta d\varphi]+H_1(r,\theta) dr+H_2(r,\theta) d\theta\} \ ,
\label{ansatzproca}
\end{equation}
for the Proca case.\cite{Herdeiro:2016tmi} The two constant parameters
$\omega,m$ are the frequency and azimuthal quantum number, with
$\omega\in \mathbb{R^+}$, $m\in\mathbb{Z}/\{0\}$. An immediate
observation is that the matter fields are not invariant under the two
aforementioned Killing vector fields:
\begin{equation}
\mathcal{L}_{\bf k} A_\mu\neq 0 \ , \qquad \mathcal{L}_{\bf m} A_\mu\neq 0 \ , \qquad \mathcal{L}_{\bf k} \Psi\neq 0 \ , \qquad \mathcal{L}_{\bf m} \Psi\neq 0 \ ,
\end{equation}
but the corresponding energy momentum-tensors are
\begin{equation}
\mathcal{L}_{\bf k}T^{(\Psi)}_{\alpha\beta}=0 \ , \qquad \mathcal{L}_{\bf k}T^{(A)}_{\alpha\beta}=0 \ , \qquad \mathcal{L}_{\bf m}T^{(\Psi)}_{\alpha\beta}=0 \ , \qquad \mathcal{L}_{\bf m}T^{(A)}_{\alpha\beta}=0 \ .
\end{equation}
Thus Bekenstein-type theorems are inapplicable and the absence of hair
is no longer guaranteed, but the left- and right-hand sides of the
Einstein equations still have the same symmetries.

The ansatz~\eqref{ansatzkbhssh}, in combination
with~\eqref{ansatzscalar} or~\eqref{ansatzproca}, yields axially
symmetric solutions. One may wonder whether BH solutions could also
exist in the much simpler spherically symmetric case, obtained by
taking $W=0$ and $F_1=F_2$ in~\eqref{ansatzkbhssh} and $m=0$ and
$\phi=\phi(r)$ in~\eqref{ansatzscalar}; $H_2=H_3=0$ and $V=V(r)$,
$H_1=H_1(r)$ in~\eqref{ansatzproca}, respectively. In that case,
however, it was shown for both the scalar case\cite{Pena:1997cy} and
the Proca case\cite{Herdeiro:2016tmi} that no BH solutions
exist. Thus, as mentioned above, symmetry (of the metric)
non-inheritance by the matter fields is a necessary but not sufficient
ingredient. A further ingredient is necessary; this can be seen by
answering question {\bf (2)} above.

\subsubsection{Stationary clouds and the threshold of superradiance}
\label{sec_clouds}
The answer to question {\bf (2)} above is ``yes.'' A test field
analysis shows the existence of stationary, everywhere regular (on and
outside the horizon) solutions of the
scalar\cite{Hod:2012px,Herdeiro:2014goa,Hod:2013zza,Benone:2014ssa} or
Proca field\cite{Herdeiro:2016tmi} on the Kerr BH spacetime:
\textit{stationary scalar or Proca clouds around Kerr BHs}. The
existence of these stationary clouds is intimately related to
superradiance, as we now illustrate for the scalar case.

The Klein-Gordon equation for a massive scalar field on the Kerr
background, $\Box_{\bf Kerr} \Psi=\mu^2\Psi$, using Boyer-Lindquist
coordinates $(t,r,\theta,\varphi)$ and an ansatz
$\Psi=e^{-i\omega t+im\varphi}S(\theta)R(r)$, allows separation of
variables and hence yields two ODEs:
\begin{equation}
\frac{1}{ \sin\theta}\frac{d}{d \theta}\left(\sin\theta \frac{d S(\theta)}{d \theta}\right)+\left[a^2\cos^2\theta(\omega^2-\mu^2)-\frac{m^2}{\sin^2\theta}+\Lambda\right]S(\theta)=0 \ ,
\label{spheroidalh}
\end{equation}
\begin{equation}
\frac{d}{d r}\left(\Delta\frac{d R(r)}{d r}\right)+\left[\frac{\omega^2(r^2+a^2)^2+a^2m^2-4Mram\omega}{\Delta}-\omega^2a^2-\mu^2r^2-\Lambda \right]R(r) =0 \ . 
\end{equation}
Here $M$ and $a$ are the ADM mass and ADM angular momentum per unit
mass of the Kerr solution, and $\Delta\equiv r^2-2Mr+a^2$. $\Lambda$ is
the separation constant, that reduces to the familiar $\ell(\ell+1)$ in
the Schwarzschild limit.

The angular equation defines the \textit{spheroidal harmonics}. To
tackle the radial wave equation, looking for bound state solutions,
one requires exponentially decaying solutions towards spatial infinity
and a purely ingoing boundary condition on the horizon (in a frame
co-rotating with the horizon). Then, one finds in general that the
frequency $\omega$ is complex:
$\omega=\mathcal{R}(\omega)+i\mathcal{I}(\omega)$. For
$\mathcal{R}(\omega)=m\Omega_H$, however, $\mathcal{I}(\omega)=0$ and
thus one finds truly stationary states with a real frequency. This
condition is interpreted as a zero mode of the superradiant
instability, which sets in for $\mathcal{R}(\omega)<m\Omega_H$
yielding $\mathcal{I}(\omega)>0$.

This bound state problem becomes particularly simple and elegant for
extremal Kerr BHs\cite{Hod:2012px}. In this case the radial equation
above, generically of confluent Heun type, reduces to the confluent
hypergeometric type, precisely the same equation one finds for the
Hydrogen atom (without spin). In this problem, the quantization
condition can be interpreted as a condition on the background
parameters. Thus, the corresponding stationary clouds -- labelled by
three quantum numbers $(n,\ell,m)$, where the first is the number of
nodes of the radial function and the last two are the spheroidal
harmonic indices -- can only exist in a subspace of Kerr solutions,
actually a one-dimensional existence line, for fixed quantum
numbers. This conclusion changes, however, when the test scalar field
is allowed to have self-interactions\cite{Herdeiro:2014pka}. The Proca
case is similar in spirit, but more involved technically, since the
Proca potentials do not decouple and no separation of variables has
been observed~\cite{Pani:2012vp,Pani:2012bp}.

To summarize: the answer to question {\bf (1)} showed that there is a
breach in the wall; the answer to {\bf (2)} shows that there
\textit{is} indeed something beyond the wall.

\subsubsection{Solitonic limits and phenomenology}
The construction of Kerr BHs with scalar and Proca hair adapted the
technology already in use for (rotating) \textit{boson
  stars}\cite{Schunck:2003kk,Liebling:2012fv}. Scalar boson stars can
be constructed with the
ansatz~\eqref{ansatzkbhssh}-\eqref{ansatzscalar} taking $r_H=0$, and
thus will be a limiting case of the corresponding Kerr BHs with scalar
hair.\footnote{To construct the scalar boson and Proca stars, it is
  useful to rescale the function $W$ as $W/r$ in~\eqref{ansatzkbhssh}
  and the function $H_1$ as $H_1/r$ in~\eqref{ansatzproca}.} The
Einstein-Klein-Gordon system of equations yields 5 coupled non-linear
PDEs for the five unknown functions plus two ``constraint" equations
(which are differentially related to the remaining ones). These
equations can be solved by a Newton-Raphson relaxation
method\cite{Herdeiro:2015gia}. Likewise, the Einstein-Proca system,
taking the ansatz~\eqref{ansatzkbhssh}-\eqref{ansatzproca}, yields 8
coupled non-linear PDEs for the eight unknown functions plus two
``constraint" equations. Solutions regular on and outside $r=r_H$ can
be found, and they correspond to Kerr BHs with Proca
hair.\cite{Herdeiro:2016tmi} The $r_H=0$ limit yields rotating
\textit{Proca stars}\cite{Brito:2015pxa}, spin-1 cousins of the
aformentioned (scalar) boson stars. These observations answer question
{\bf (3)} above.

The exploration of the physical and phenomenological properties of
these new families of hairy BHs connected to the Kerr solutions is
ongoing research. For the scalar case it has been noted that the hairy
BHs can have quadrupoles and orbital frequency at the ISCO quite
distinct from the Kerr
case\cite{Herdeiro:2014goa,Herdeiro:2015gia}. Particularly striking
are the BH shadows that have been obtained for some examples of Kerr
BHs with scalar hair, with remarkably different shapes and sizes from
the Kerr case\cite{Cunha:2015yba}. These shadows can be partly
understood regarding the hairy BHs as composites of a boson star with
a horizon, a perspective that can also explain, for instance, the
ergoregion structure of these
spacetimes\cite{Herdeiro:2014jaa}. Generalizations of the hairy BHs to
include self-interactions of the matter field have been
considered\cite{Kleihaus:2015iea,Herdeiro:2015tia}. It is likely that
similar generalizations are possible in the Proca case.

Finally, let us remark that \textit{Myers-Perry BHs with scalar hair}
have been found in $D=5$. These are also anchored to a similar
condition between the frequency of the scalar field and the angular
velocity of the horizon.\cite{Brihaye:2014nba,Herdeiro:2015kha} In
$D=5$ asymptotically flat spacetimes, vacuum Myers-Perry BHs are not,
however, afflicted by superradiant instabilities of massive scalar
fields. As such, when the scalar field is set equal to zero, the hairy
solutions do not reduce to vacuum Myers-Perry solutions: even though
the local geometry can become arbitrarily close to that of the vacuum
solutions, there is always a \textit{mass gap}. A generalization of
these solutions including higher curvature terms has also been
constructed.\cite{Brihaye:2015qtu}

\subsubsection{Can hairy BHs form?}
\label{hairyform}
The existence of Kerr BHs with scalar and Proca hair is theoretically
interesting, and it presents us with a rich landscape of previously
unknown BH solutions in GR. These solutions require complex bosonic
fields, but extremely long-lived solutions exist even for {\it real}
fields. These solutions describe BHs surrounded by a ``cloud'' of
massive bosons~\cite{Okawa:2014nda}.  Are these solutions relevant for
astrophysics? The answer to this question depends on two main issues:
(A) the existence of massive (and very light) bosonic fields in
Nature, and (B) the formation mechanism of these solutions and their
stability properties. 

Question (A) is an open issue. Question (B) has been studied in a
specific scenario. The development of the superradiant instability of
massive scalars and vectors has recently been addressed taking into
account gravitational radiation, superradiant growth and the effects
of a putative accretion disk around the BH~\cite{Brito:2014wla}, but
in an adiabatic approximation (rather than a fully non-linear
numerical evolution).  Assuming that the bosonic cloud is formed
through the development of the superradiant instability, it was shown
that, within the previous approximations, (i) the observation of
supermassive BHs would show gaps in the Regge-plane, corresponding to
BHs which quickly become unstable due to superradiant effects; (ii)
the bosonic cloud {\it never} backreacts significantly on the
geometry; and even though a hairy BH can effectively form, it does not
depart significantly from the Kerr geometry~\cite{Brito:2014wla}.

Progress on question (B) has also been achieved using a different toy
model: a Reissner-Nordstr\"om BH enclosed in a cavity. This system is
afflicted by the superradiant instability of bosonic fields (not
necessarily massive, since the trapping mechanism is now provided by
the cavity) and it was observed that superradiant instabilities in
this system -- at the test-field level -- grow much faster than for
Kerr BHs, occurring even for spherically symmetric
modes\cite{Herdeiro:2013pia,Hod:2013fvl,Degollado:2013bha,Hod:2016nsr}. These
two features make the system tractable with current numerical
relativity technology, allowing us to perform fully non-linear
evolutions of the superradiant
instability~\cite{Sanchis-Gual:2015lje}. The simulations showed that
the final states of these unstable BHs are indeed hairy BHs at the
threshold of superradiance, which can be regarded as the charged
counterparts (in this context) of Kerr BHs with scalar
hair~\cite{Dolan:2015dha}. Similar results have also been obtained for
superradiantly unstable charged BHs in anti-de-Sitter
spacetime\cite{Bosch:2016vcp}.

Finally, an orthogonal process for the formation of these hairy BHs
could be starting from the solitonic limit, rather than the non-hairy
BH limit. It would be very interesting to understand if unstable
rotating scalar boson (or Proca) stars could develop into hairy BHs,
and how hairy these BHs would be. This is an open issue.

\section{Analog gravity}
\label{sec_analog}

Analog models of gravity are a useful tool to investigate kinematical aspects 
of curved spacetimes in condensed matter systems.~\cite{Unruh:1980cg, Visser:1997ux}
Analogues have been presented in many contexts, like Bose-Einstein condensates, 
optical media, and fluids.~\cite{Novello:2002qg, Barcelo:2005fc, AMproc}
Here we will give emphasis to the progress made in the latter context.
Indeed, many interesting physical properties of sonic analogues of BHs have 
been recently studied, like, for instance, absorption and scattering 
phenomena,~\cite{Basak:2002aw, Crispino:2007zz, Oliveira:2010zzb, Dolan:2009zza, Dolan:2011zza, Dolan:2012yc}
as well as quasinormal modes (QNMs).~\cite{Berti:2004ju, Cardoso:2004fi, Dolan:2010zza, Dolan:2011ti}

\subsection{Acoustic analogues}
Propagation of sound waves in an ideal fluid, under certain considerations, 
may be described using the Klein-Gordon equation
for a massless scalar field $\Phi$ in an effective curved spacetime, namely
\begin{equation}
 \Box \Phi = \frac{1}{\sqrt{|g|}} \partial_\mu \left( \sqrt{|g|} g^{\mu \nu} \partial_\nu \Phi \right) = 0 , 
 \label{kg}
\end{equation}
where $g_{\mu \nu}$ are the covariant components of the effective metric 
($g^{\mu \nu}$ being its contravariant components), 
with determinant $g$. 
We should emphasize that $g_{\mu \nu}$ is a function of the local properties of the fluid and, 
in general, it is not a solution of Einstein's equations.

\subsubsection{Canonical acoustic hole}
The simplest acoustic analogue to a BH is the so-called ``canonical acoustic hole''.
It consists of a spherically symmetric steady flow of an irrotational barotropic fluid
(considered also as incompressible and inviscid),
presenting a sink at the origin.
It may be described by the following line element:
\begin{equation}
ds_{\text{cah}}^{2}=-f(r) \, c_s^2 dt^{2}+\left[ f(r) \right]^{-1}dr^{2}+r^{2} \left( d
\theta^{2}+\sin^{2}\theta d\phi^{2} \right).
\label{cah}
\end{equation}
Here 
\begin{equation}
f(r)\equiv 1-r_{H}^{4}/r^{4}, 
\label{f}
\end{equation}
where $r_{H}$ is
the radius of the sonic event horizon, inside which 
the radial velocity exceeds the speed of sound $c_s$ in the fluid. 
The canonical acoustic hole is an analogue of the Schwarzschild BH.

\subsubsection{Draining bathtub}
An acoustic analogue of a rotating BH is the so-called ``draining bathtub",
whose line element may be written as
\begin{equation}
ds_{\text{dbt}}^{2}=-h(r) \, c_s^2 dt^{2}+\left[ h(r) \right]^{-1}dr^{2} + 
\left( r d\phi - Cdt/r \right)^{2}.
\label{dbt}
\end{equation}
Here
\begin{equation}
h(r)\equiv 1-\frac{D^{2}}{c_s^2 r^{2}}, 
\label{h}
\end{equation}
and the constants $C$ and $D>0$ stand for the circulation and the draining, respectively.
This effective geometry corresponds to the one experienced by sound waves 
propagating in a fluid with flow velocity
\begin{equation}
\mathbf{v}
= - \frac{D}{r}  \hat{r} + \frac{C}{r}  \hat{\phi}\ .
\label{vdbt}
\end{equation}
The draining bathtub has an ergoregion 
(defined by the supersonic flow condition $|\mathbf{v}| 
\ge c_s$) within the radius $r^{\text{dbt}}_{e} = \sqrt{C^2+D^2}/c_s$, 
and a sonic horizon 
(defined by $\mathbf{v} \cdot \hat{r} = c_s$) at radius
$r^{\text{dbt}}_H= D/c_s$.~\cite{Oliveira:2010zzb}

\subsubsection{Hydrodynamic vortex}
By setting $D=0$ in Eq.~(\ref{vdbt}), we are left with a purely circulating flow, 
which, in a (3+1)-dimensional setup, may be associated to the following line element
\begin{equation}
ds_{\text{hv}}^2=-\left(1-\frac{C^2}{c_s^2r^2} \right) c_s^2 dt^2+dr^2-2Cdt d\phi+r^2d\phi^2+dz^2\,,
\label{vortex}
\end{equation}
where $r^{\text{hv}}_e \equiv |C|/c_s$ is the outer boundary of the ergoregion.
This effective spacetime is the so-called ``hydrodynamic vortex".

In the remainder of this section we will set $c_s\equiv1$.

\subsection{Ergoregion instabilities in acoustic systems}
Ergoregion instabilities in acoustic systems have been recently
studied for the hydrodynamic vortex, both for
incompressible~\cite{Oliveira:2014oja} and for compressible
fluids~\cite{Oliveira:2015vqa}.  Here we will review the investigation
of instabilities of the hydrodynamic vortex composed by an
incompressible fluid~\cite{Oliveira:2014oja}.  (The
numerical results exhibited here are obtained for higher values of the
azimuthal number $m$, complementing the ones exhibited in
Ref.~\refcite{Oliveira:2014oja}).

Using the line element~\eqref{vortex} in the Klein-Gordon
equation~\eqref{kg}, and assuming a decomposition of the field $\Phi$
as
\begin{equation}
\Phi(t,r,\phi, z) = \frac{1}{\sqrt{r}} \sum_{m=-\infty}^{\infty} u_{\omega m}(r) \exp\left[ {i \left(m \phi-\omega t\right)}\right],
\label{Phi_1}
\end{equation}
we find the ordinary differential equation
\begin{eqnarray}
\left[\frac{d^2}{dr^2}+\left( \omega-\frac{Cm}{r^2}\right)^2-V^{\text{hv}}_m(r)\right]u_{\omega m}(r)=0\,,
\label{radial}
\end{eqnarray} 
where the effective potential $V^{\text{hv}}_m(r)$ is given by
\begin{equation}
V^{\text{hv}}_m(r) = \frac{m^2 - 1/4}{r^2},
\end{equation}
where $m$ is an integer number related with the angular momentum, and
$\omega$ is the frequency of the perturbation.

Solutions describing QNMs can be obtained from Eq.~\eqref{radial}, 
considering the asymptotic behavior at large radial distances
\begin{equation}
u_{\omega m}\left(r\to \infty \right) \sim \exp\left( {i\omega r}\right)\,,
\label{Boun_1}
\end{equation}
and a boundary condition of Neumann type at $r=r_{\rm min}$ (close to
the center of the vortex):
\begin{equation}
\left[\frac{d}{dr}\left( \frac{u_{\omega m}(r)}{\sqrt{r}}\right) \right]_{r=r_{\rm min}}=0\,.
\label{Boun_2}
\end{equation}
This condition is related to a cutoff on the radial velocity increment, i.e., $\left[\frac{\partial \Phi}{\partial r}\right]_{r=r_{\rm min}}=0$ [cf. Eq.~\eqref{Phi_1}]~\cite{Oliveira:2014oja}.

\begin{figure}[hbtp!]
\includegraphics[width=0.5\textwidth]{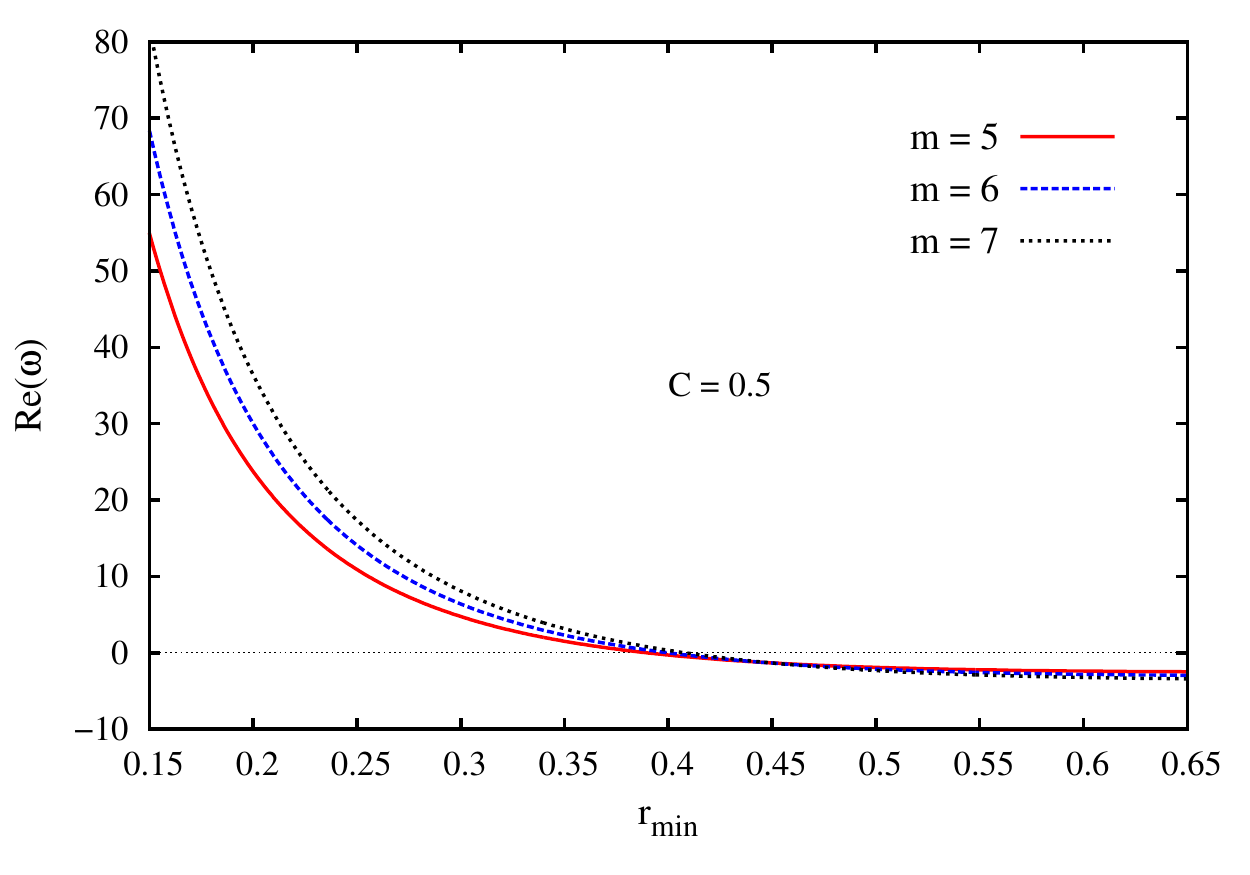}\includegraphics[width=0.5\textwidth]{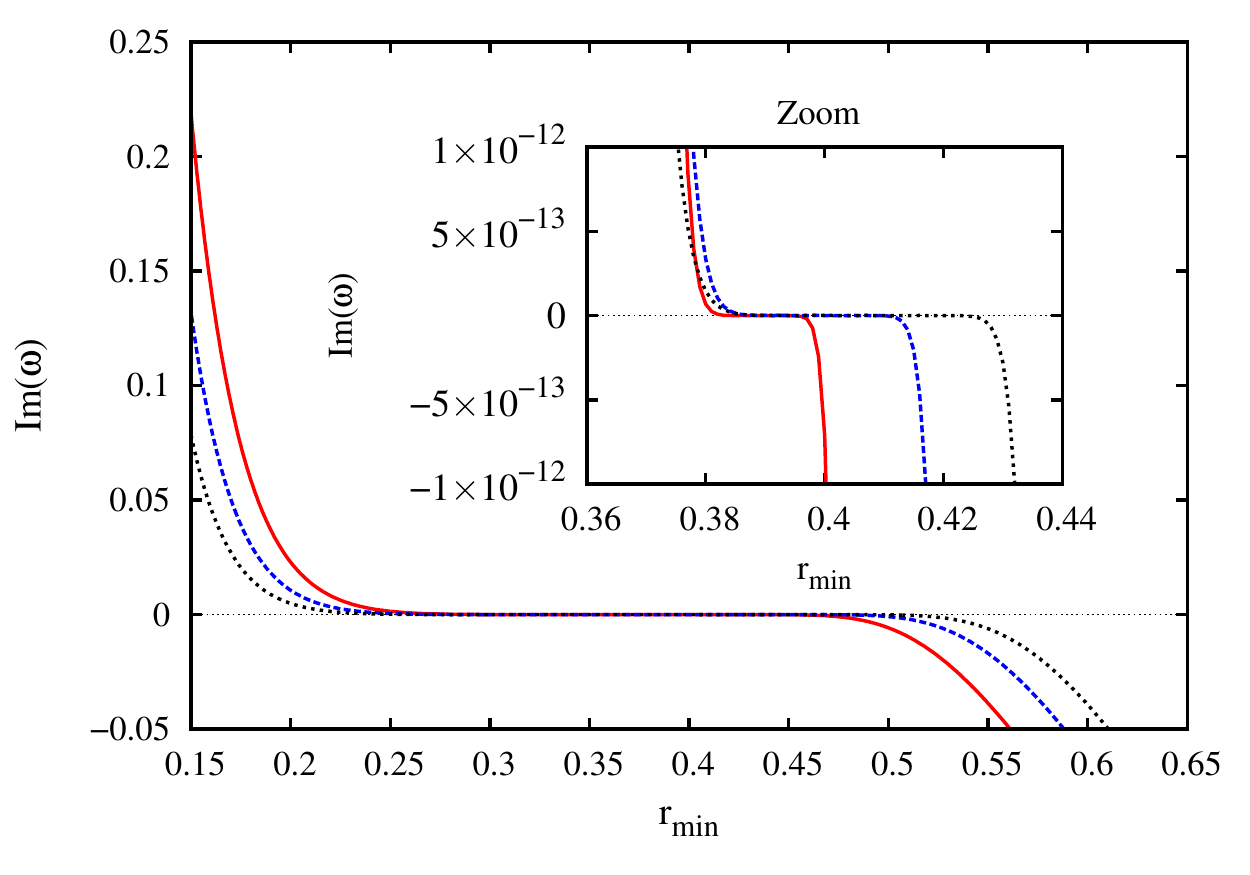}
\caption{Real (left) and imaginary (right) components of the
  fundamental QNM frequencies, plotted as a function of $r_{\rm min}$,
  for $C=0.5$ and $m=5,\,6,\,7$. These results were obtained via the
  CF method.  }
\label{fig_vortex}
\end{figure}

Ergoregion instabilities are present in acoustic systems possessing an
ergoregion but not an event horizon.  These instabilities are
developed inside the ergoregion, i.e., for
$r < |C| $~\cite{Oliveira:2014oja}.  In order to obtain the QNM
frequencies of the hydrodynamic vortex, we can use different numerical
techniques to integrate Eq.~\eqref{radial} in the frequency domain.
Some results for QNM frequencies are exhibited in
Table~\ref{table_vortex}, for different values of the azimuthal number
$m$ and $r_{\rm min}$, obtained using two different frequency-domain
methods, namely the direct integration (DI) and continued fraction
(CF) methods.
\begin{table}
\tbl{QNM frequencies $\omega$ for different values of the azimuthal number $m$ and circulation $C=0.5$, 
obtained numerically from estimates via the DI and CF methods. 
We impose the asymptotic behavior given by Eq.~\eqref{Boun_1} 
and a boundary condition of Neumann type, represented by Eq.~\eqref{Boun_2}, 
at $r_{\rm min}=0.51$ (outside the ergoregion) and $r_{\rm min}=0.25$ (inside the ergoregion).}
{\scalebox{1.2}{\begin{tabular}{@{}c c c c c c@{}}
\hline
 \multicolumn{1}{c}{} & \multicolumn{1}{c}{}& \multicolumn{2}{c}{$r_{\rm min}=0.51$} & \multicolumn{2}{c}{$r_{\rm min}=0.25$} \\
\hline
\hline
 \multicolumn{1}{c}{$m$} & \multicolumn{1}{c}{Method}& \multicolumn{1}{c}{$\text{Re}(\omega) $} & \multicolumn{1}{c}{$\text{Im}(\omega) $} & \multicolumn{1}{c}{$\text{Re}(\omega) $} & \multicolumn{1}{c}{$\text{Im}(\omega) $} \\
\hline
\multirow{2}{*}{$ 5 $} & DI & $-1.98856262$ & $-0.00968749$ & $+10.90342057  $ & $+0.00145905   $ \\
                       & CF & $-1.98856262$ & $-0.00968749$ & $+10.90342057  $ & $+0.00145905   $  \\
\hline
\multirow{2}{*}{$ 6 $} & DI & $-2.24470575$ & $-0.00198696$ & $+14.09001520  $ & $+0.00050399  $ \\
                       & CF & $-2.24470575$ & $-0.00198696$ & $+14.09001520  $ & $+0.00050399  $ \\
\hline
\multirow{2}{*}{$ 7 $} & DI & $-2.47088637$ & $-0.00029125$ & $+17.36697101  $ & $+0.00017138   $\\
                       & CF & $-2.47088637$ & $-0.00029125$ & $+17.36697101  $ & $+0.00017138   $ \\
\hline\hline
\end{tabular}\label{table_vortex}}
}
\end{table}  
Real and imaginary parts of the QNM frequencies are plotted in Fig.~\ref{fig_vortex}, as functions of $r_{\rm min}$, 
for different values of the azimuthal number $m$, obtained using the CF method, considering a circulation $C=0.5$. 

From the results exhibited in Table~\ref{table_vortex}, it can be seen
that as the azimuthal number $m$ increases, the magnitude of the real
(imaginary) part of the QNM frequencies increases
(decreases). Moreover, from the plots exhibited in
Fig.~\ref{fig_vortex}, we find that as $r_{\rm min}$ decreases, the
magnitude of the real and imaginary parts of the QNM frequencies
increase (decrease) for unstable (stable) modes. This behavior of the
imaginary part can be clearly seen in the inset of the right panel of
Fig.~\ref{fig_vortex}.

\subsection{Acoustic clouds}
As analogues to the clouds around rotating BHs, described in Section~\ref{sec_clouds}, we may have 
acoustic clouds around the draining bathtub.
Taking advantage of the symmetries of 
the draining bathtub spacetime, characterized by Eq.~(\ref{dbt}),
we can search for solutions of the Klein-Gordon equation (\ref{kg}), 
assuming the separation of variables
\begin{equation}
\Phi_m(r,\phi,t) = e^{i(m \phi-\omega t)}\zeta_m(r).
\end{equation}
The radial function $\zeta_m(r)$ obeys the ordinary differential equation
\begin{equation}
\frac{h(r)}{r}
\frac{d}{dr}\left[r h(r) \frac{d\zeta_m}{dr}\right] + 
\left[\omega^2-\frac{2Cm\omega}{r^2}-\frac{m^2}{r^2}\left(1-\frac{D^2+C^2}{r^2}\right)\right]\zeta_m = 0.
\label{era}
\end{equation}
Using the tortoise coordinate, defined by
\begin{equation}
\frac{d}{dr_*} \equiv h(r) \frac{d}{dr}\,,
\end{equation}
we can rewrite Eq.~(\ref{era}) as the Schr\"odinger-like equation
\begin{equation}
\frac{d^2}{dr_*^2}u_m + \left[\left(\omega - \frac{C m}{r^2}\right)^2 - V^{\text{dbt}}_m(r)\right]u_m = 0\ ,
\label{era2}
\end{equation}
where $u_m \equiv \sqrt{r}\zeta_m$, and we have defined the effective potential
\begin{equation}
V^{\text{dbt}}_m(r) = h(r) \left[\frac{m^2-1/4}{r^2}+\frac{5 D^2}{4 r^2}\right] \ .
\end{equation}

Considering the asymptotic limit of Eq. (\ref{era2}), we find the solutions
\begin{equation}
\zeta_{\omega m}(r) \sim 
\left\{ 
\begin{array}{ll}
e^{-i(\omega-\omega_c) r_*}, \quad &\mbox{for $r\rightarrow r_H$}\ ,\\
 e^{i\omega r_*}, \quad &\mbox{for $r\rightarrow \infty$}\ .
\end{array}
\right.
\label{sol}
\end{equation}
In order to have clouds we must choose $\omega=\omega_c \equiv m \Omega_H$,
and enclose the system inside a ``barrier'' located at $r=r_0$.
At the ``barrier'' we impose suitable boundary conditions,
usually chosen to be of Dirichlet or Neumann type.~\cite{Oliveira:2014oja}

\begin{figure}[htpb!]
\begin{center}
\includegraphics[width=2.6in]{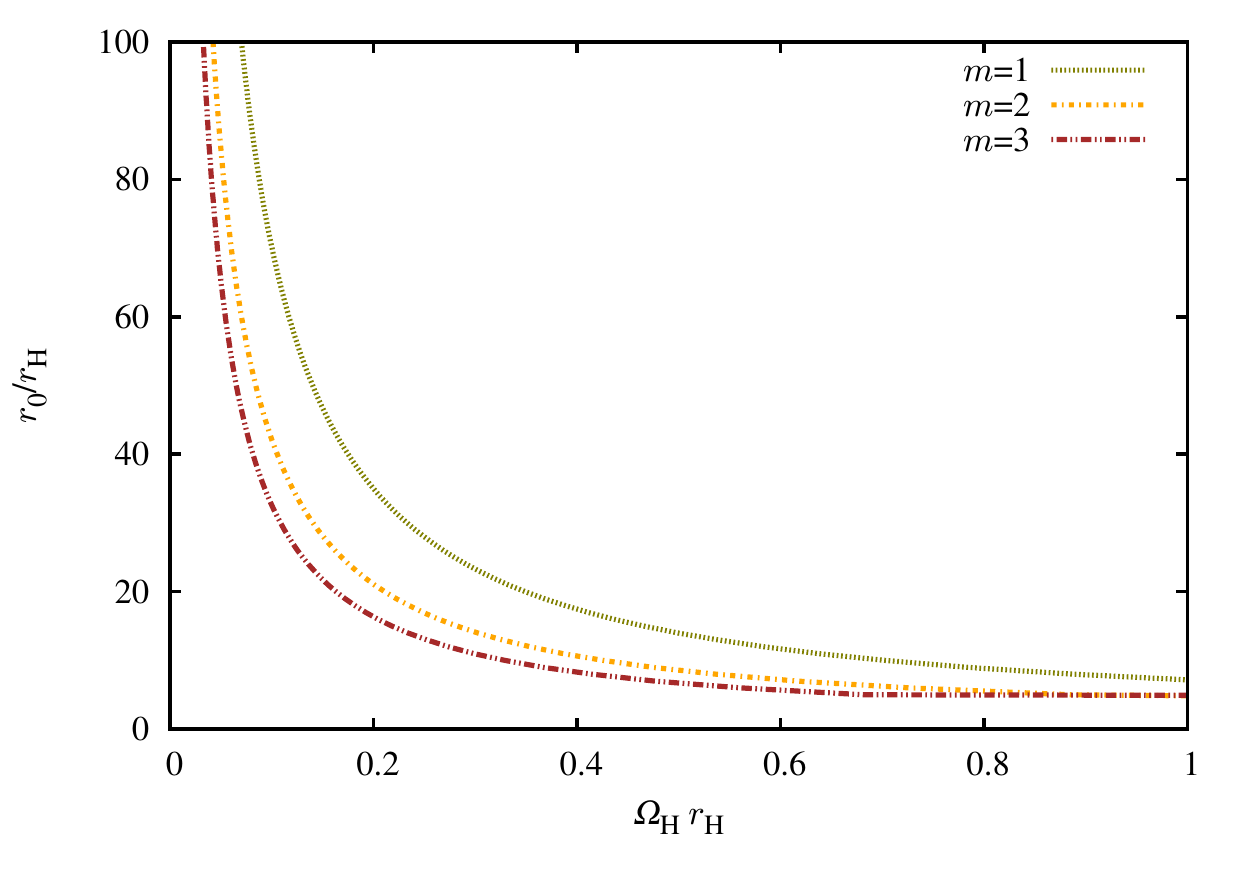}\includegraphics[width=2.6in]{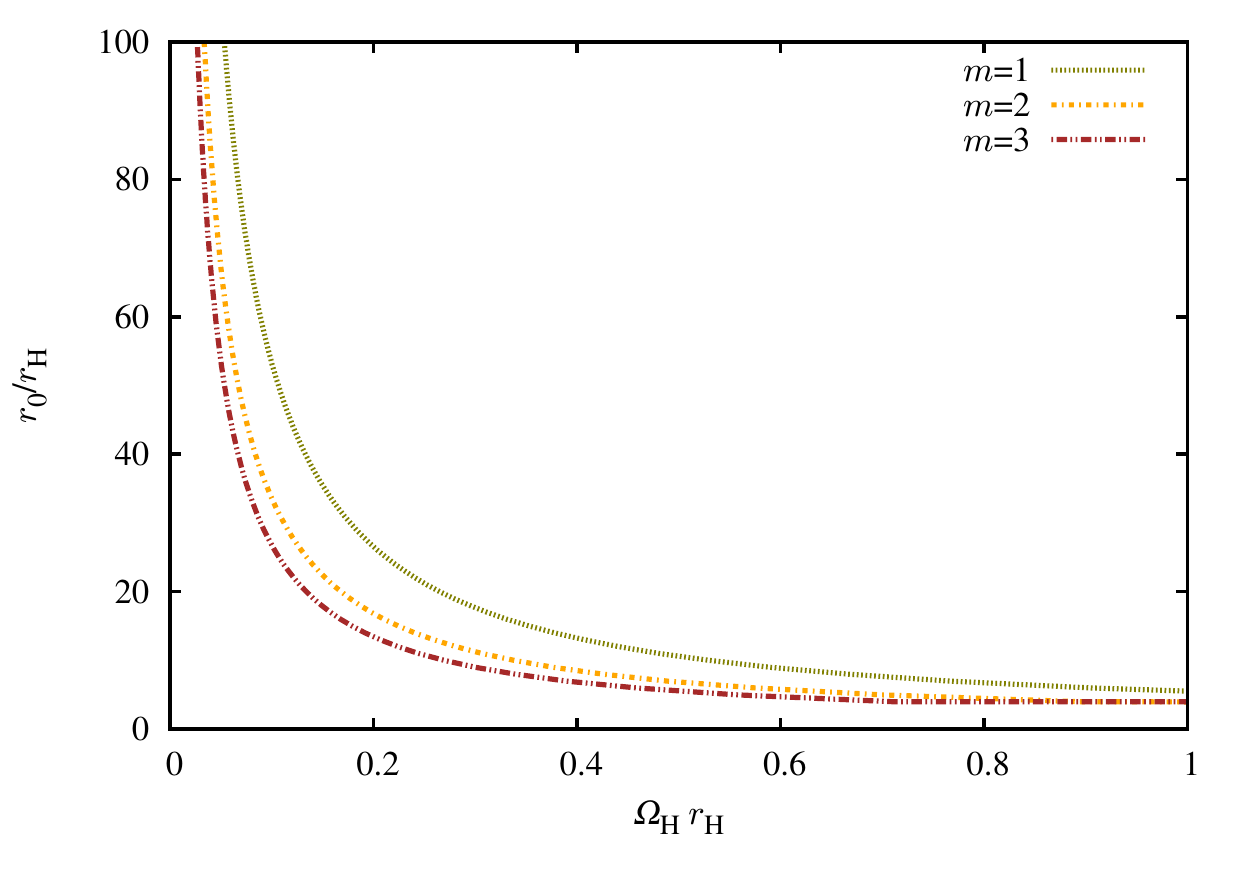}
\end{center}
\caption{Acoustic clouds around the draining bathtub surrounded by a
  boundary at $r=r_0$, with angular velocity at the horizon
  $\Omega_H$, for $n=1$ and different values of $m$, for Dirichlet
  (left panel) and Neumann (right panel) boundary conditions.  The
  number $n$ denotes the node number of the radial function.  Similar
  figures, but with different choices of the clouds quantum numbers
  can be found in Benone et al.~\cite{Benone:2014nla, Benone:2015jda}}
\label{tacs}
\end{figure}

In Fig.~\ref{tacs} we analyze the behavior of the acoustic clouds by
plotting the values of the frontier location $r_0$ as a function of
the angular velocity at the horizon $\Omega_H$, for different choices
of the azimuthal number $m$.  We see that, for a fixed position $r_0$
of the barrier, the acoustic clouds occur for smaller values of
$\Omega_H$ as we increase the value of $m$.

\begin{figure}[htpb!]
\begin{center}
\includegraphics[width=2.6in]{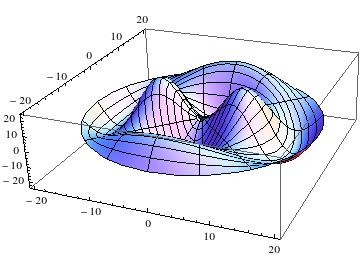}\includegraphics[width=2.6in]{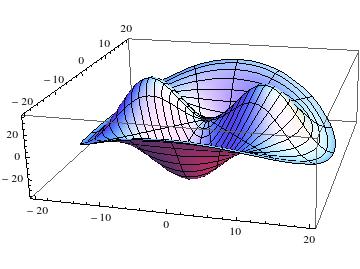}
\end{center}
\caption{Real part of $\Phi$ in the $x,y$ plane, for $r_0/r_H=20$,
  $n=1$ and $m=2$. The left panel displays the case for Dirichlet
  boundary conditions, with $C/r_H=0.21$, while the right panel
  displays the case for Neumann boundary conditions, with
  $C/r_H=0.17$.}
\label{tacs1}
\end{figure}

Three-dimensional plots of the radial and azimuthal profiles of
acoustic clouds are shown in Fig.~\ref{tacs1}, for Dirichlet (left
panel) and Neumann (right panel) boundary conditions.

\clearpage

\section{Concluding Remarks}
\label{sec:conclusions}
The fantastic conceptual and formal elegance of Einstein's gravity
hides a tremendous complexity when it is applied to realistic,
dynamical systems. Quite often, all hope of finding elegant analytic
solutions is lost. Then, to tackle this complexity, one needs to
resort to numerical solutions. This necessity is now well understood
by the scientific community and with the current available techniques,
together with the ones under development, there is a strong belief
that a lot can be learned about the most elegant physical theory -- or
generalizations thereof -- in the strong field, dynamical regime. We
foresee interesting times ahead.

\section*{Acknowledgments}
We would like to thank all the participants in the NRHEP network and all colleagues that have collaborated in the works reported in this review.
E.B. was supported by NSF CAREER Grant No. PHY-1055103 and by FCT
contract IF/00797/2014/CP1214/CT0012 under the IF2014 Programme.
V.C. and U.S. acknowledge financial support provided under the
European Union's H2020 ERC Consolidator Grant ``Matter and
strong-field gravity: New frontiers in Einstein's theory'' grant
agreement no. MaGRaTh--646597.
L.C. thanks Carolina L. Benone and Leandro A. Oliveira for useful discussions, as well as
Conselho Nacional de Desenvolvimento Cient\'\i fico e Tecnol\'ogico (CNPq),
Coordena\c{c}\~ao de Aperfei\c{c}oamento de Pessoal de N\'\i vel Superior (CAPES) and
Funda\c{c}\~ao Amaz\^onia de Amparo a Estudos e Pesquisas do Par\'a (FAPESPA)
for partial financial support.
C.H. acknowledges
funding  from  the  FCT-IF  programme.   
Research at Perimeter Institute is supported by the Government of
Canada through Industry Canada and by the Province of Ontario through
the Ministry of Economic Development $\&$ Innovation.
This work was supported by the H2020-MSCA-RISE-2015 Grant No.
StronGrHEP-690904,
the CIDMA project UID/MAT/04106/2013,
STFC Consolidator Grant No. ST/L000636/1,
the SDSC Comet and TACC Stampede clusters through NSF-XSEDE Award
No.~PHY-090003, the Cambridge High Performance
Computing Service Supercomputer Darwin using Strategic Research
Infrastructure Funding from the HEFCE and the STFC, and DiRAC's Cosmos
Shared Memory system through BIS Grant No.~ST/J005673/1 and STFC Grant
Nos.~ST/H008586/1, ST/K00333X/1.
%
%

\newpage 

\bibliographystyle{ws-ijmpd}
\bibliography{biblio}

\end{document}